\documentclass[preprint,12pt,authoryear]{elsarticle}

\usepackage{amssymb}
\usepackage{xcolor}
\usepackage{amsmath}
\usepackage{booktabs}

\newtheorem{teo}{Theorem}
\newtheorem{lema}[teo]{Lemma}

\journal{New Astronomy Reviews}

\begin{document}

\begin{frontmatter}



\title{A review on fundamental bounds and estimators for photometry and astrometry of celestial point sources using array detectors, from first principles} 


\author[label1]{Sebastián Espinosa} 
\author[label2]{Rene A. Mendez}
\author[label3]{Jorge F. Silva}
\author[label3]{Marcos Orchard}
\affiliation[label1]{organization={Universidad Técnica Federico Santa María, Advanced Center for Electrical and Electronic Engineering},
 addressline={General Bari 699}, 
           city={Valparaiso},
           postcode={2390136}, 
           country={Chile}}
\affiliation[label2]{organization={Department of Astronomy, Universidad de Chile},
           addressline={Beauchef 850}, 
          city={Santiago},
           postcode={8370448}, 
           country={Chile}}
 \affiliation[label3]{organization={Department of Electrical Engineering, Universidad de Chile},
          addressline={Beauchef 850}, 
          city={Santiago},
            postcode={8370451}, 
            country={Chile}}

\begin{abstract}
Precise astrometric and photometric measurements of celestial point sources are fundamental to modern astronomy. These measurements, used to determine object positions, motions, and fluxes, are based on observational models that have evolved from empirical centroiding rules to rigorous probabilistic formulations at the pixel level. This review summarizes key contributions that formalized this transition and analyzes seminal works addressing both the theoretical limits and the empirical performance of estimators. Central to these developments is the derivation of fundamental bounds, such as the Cramér-Rao Lower Bound (CRLB), and the assessment of widely used estimators, including Maximum Likelihood (ML), Least Squares (LS), and Weighted Least Squares (WLS). These studies show that, while the CRLB sets a theoretical benchmark, practical estimators achieve it only under specific signal-to-noise ratio (SNR) regimes, with notable discrepancies in high-SNR conditions. Moreover, recent results demonstrate that jointly estimating source flux and background significantly improves photometric precision compared to sequential approaches. Looking ahead, the increasing complexity of astronomical surveys, driven by massive data volumes, dynamic observational conditions, and the integration of machine learning, poses new challenges to reliable inference. In this context, tools from statistical theory, including performance bounds and theoretically grounded estimators, remain critical to guide algorithm design and ensure robust astrometric and photometric pipelines.
\end{abstract}


\begin{highlights}
\item We review fundamental limits and estimator performance for astrometry and photometry of point sources, from first principles under Poisson noise models.

\item We analyze the interplay between signal-to-noise ratio, pixel sampling and summarize key results using Cramér-Rao bounds and classical estimators.

\end{highlights}

\begin{keyword}
Astronomical instrumentation, Stellar photometry, Estimation theory, Fisher information, Likelihood estimation



\end{keyword}

\end{frontmatter}



\section{Introduction}
Astrometry and photometry are foundational tools in observational astronomy, enabling the precise measurement of celestial object positions, proper motions, distances, and fluxes. These measurements support major scientific advances, from the detection of exoplanets and the identification of stellar populations, to the reconstruction of the Milky Way's structure~\citep{perryman1997hipparcos, gaia2016}. Attaining high precision in these domains hinges not only on technological advances, but also on a rigorous understanding of the statistical structure of the observational process, including its fundamental limits and the performance of the algorithms employed.

This need for precision has deep historical roots. Astrometry, one of the oldest branches of astronomy, began with efforts to chart the night sky, with Hipparchus of Nicaea (b. ca 190 BC) among the first to systematically catalogue stellar positions~\citep{hoskin1997hipparchus}. Over centuries, it has evolved from geometric triangulation methods to space-based missions such as Hipparcos and Gaia, which have brought astrometric precision to unprecedented levels~\citep{mignard2018gaia, 2024Ap&SS.369...23H}. Similar progress has occurred in photometry through the emergence of massive imaging surveys such as SDSS, Pan-STARRS, LSST, JWST, and Euclid, each of which has pushed the limits of source modeling, calibration, and error characterization. These advances were enabled not only by improved instrumentation, but also by increasingly sophisticated mathematical models and statistical techniques. In modern astrometry and photometry, precision depends critically on the probabilistic modeling of image formation and noise, and on the design of estimators that approach the fundamental limits imposed by information theory~\citep{sozzetti2005}.

The statistical challenges underlying such precision are nontrivial. Modern digital detectors and photon-counting methods yield data governed by Poisson statistics and spatially structured noise, requiring sophisticated probabilistic modeling~\citep{adorf1996limits}. In this context, estimation theory—and in particular, the Cramér-Rao Lower Bound (CRLB)—plays a central role. The CRLB establishes a theoretical lower bound on the variance of any unbiased estimator and defines the best achievable precision for position and flux estimation in a given observational model~\citep{rao1945, kay1993fundamentals}.

The CRLB is computed from the Fisher information matrix (FIM), which quantifies the sensitivity of the likelihood function to changes in the parameters of interest. This connection links astronomy with a rich statistical tradition, where maximum likelihood estimation (ML), least-squares (LS), and their variants serve as the primary tools for inference. These methods have been widely adopted in astronomical data reduction pipelines due to their balance of interpretability and asymptotic efficiency~\citep{lupton1993statistics}.

Yet, practical estimators only achieve the CRLB under idealized conditions. In realistic scenarios, their performance varies with signal-to-noise ratio (SNR), point spread function (PSF) structure, background noise, and detector resolution \citep{liaudat2023point,xin2018study}. In response, several extensions have emerged: weighted least-squares (WLS) and adaptive estimators address varying noise conditions~\citep{espinosa2018optimality}, while the Bayesian Cramér–Rao Lower Bound (BCRLB) incorporates prior information to improve estimation in low-SNR regimes~\citep{echeverria2016analysis}.

More recently, statistical approaches have also tackled emerging challenges in astronomy, such as motion blur in fast-moving objects~\citep{bouquillon2017characterizing} and satellite streak contamination~\citep{lawrence2022}. These developments highlight a broader trend: astronomy is increasingly reliant on statistical theory and modeling not only for data processing but also for framing fundamental limits and guiding optimal measurement design.

In this review, we systematically examine the theoretical and empirical performance of estimators used in astrometry and photometry. We discuss foundational contributions on the CRLB, explore their implications across one- and two-dimensional settings, and evaluate the behavior of classical and modern estimators under realistic observational conditions. Our goal is to offer a unified perspective on precision estimation from first principles, bridging statistical theory with contemporary astronomical practice.

The remainder of this review is organized as follows. Section~\ref{sec:model} introduces the observational model that supports astrometric and photometric measurements, emphasizing its historical development and modern probabilistic formulation. Section~\ref{sec:bounds} presents fundamental performance limits, with a focus on the CRLB and its relevance in assessing estimator precision. Section~\ref{sec:astro_estimators} reviews the performance of classical and modern estimators in astrometry, while Section~\ref{sec:photo_estimators} analyzes their behavior in photometric tasks, including flux and background estimation. Section~\ref{sec:extensions} examines how Fisher information and related bounds, extend to more general settings—including Bayesian inference, moving-source models, and applications in modern photometric surveys. Finally, Section~\ref{sec:conclusions} summarizes the main insights and outlines open directions for future research at the intersection of statistical theory and observational astronomy.

\section{Modeling evolution in astrometry and photometry}
\label{sec:model} 
\subsection{Foundations of the observational model}

The observational model used in astrometry and photometry has evolved significantly over the past decades, driven by both theoretical developments and improvements in instrumentation. Foundational efforts in astrometry date back to the late 20th century, with projects such as the \textit{Carte du Ciel} and early photographic surveys like the Bonner Durchmusterung and Cape Photographic Durchmusterung~\citep{seidelmann1992explanatory,walter2000astrometry}. These initiatives developed systematic methods for position measurements and reduction techniques using photographic plates, laying the groundwork for later statistical models.

Later, in the 1980s, \citet{vanaltena1986} introduced refined models for photographic astrometry that treated stellar centroids using empirical rules and Gaussian fitting over simplified PSF approximations. While these models advanced centroid precision, they were still limited by systematic errors and a relatively poor understanding of the detector response that affected photographic plates.

Subsequently, works by \citet{lindegren1978,lindegren2008} formalized a rigorous statistical framework for modeling the image-formation process, incorporating the instrument's PSF, calibration uncertainties, and time-dependent effects. Building on these principles, pixel-level modeling for Gaia was later developed in \citet{prod2012impact,crowley2016gaia}, where the PSF/LSF structure, detector effects, and noise statistics were explicitly linked to the astrometric solution. These developments established the practical foundations for unbiased centroid determination and rigorous error propagation in large-scale space missions.

In parallel, Hans-Martin Adorf and colleagues helped advance the statistical treatment of photometric measurements by introducing maximum likelihood methods~\citep{adorf1995,adorf1996limits}. They explored how flux estimators behave under Poisson noise and showed the shortcomings of traditional aperture photometry, especially in crowded fields. Their work encouraged the adoption of PSF fitting techniques to achieve better accuracy in complex imaging conditions.

With the transition from photographic plates to digital detectors such as charge-coupled devices (CCDs), modern models now treat each pixel as a Poisson-distributed random variable whose expectation combines the source signal modulated by the PSF and a background term. This pixel-based statistical modeling has enabled the derivation of fundamental performance bounds (e.g., via the FIM) and the development of optimal estimators under varying observational regimes. In the next subsection, we present the mathematical formulation of this model, focusing on how photon counts are modeled statistically and how key components interact within this framework.

\subsection{Mathematical formulation}
\begin{figure}[ht]
    \centering
    \includegraphics[width=0.65\textwidth]{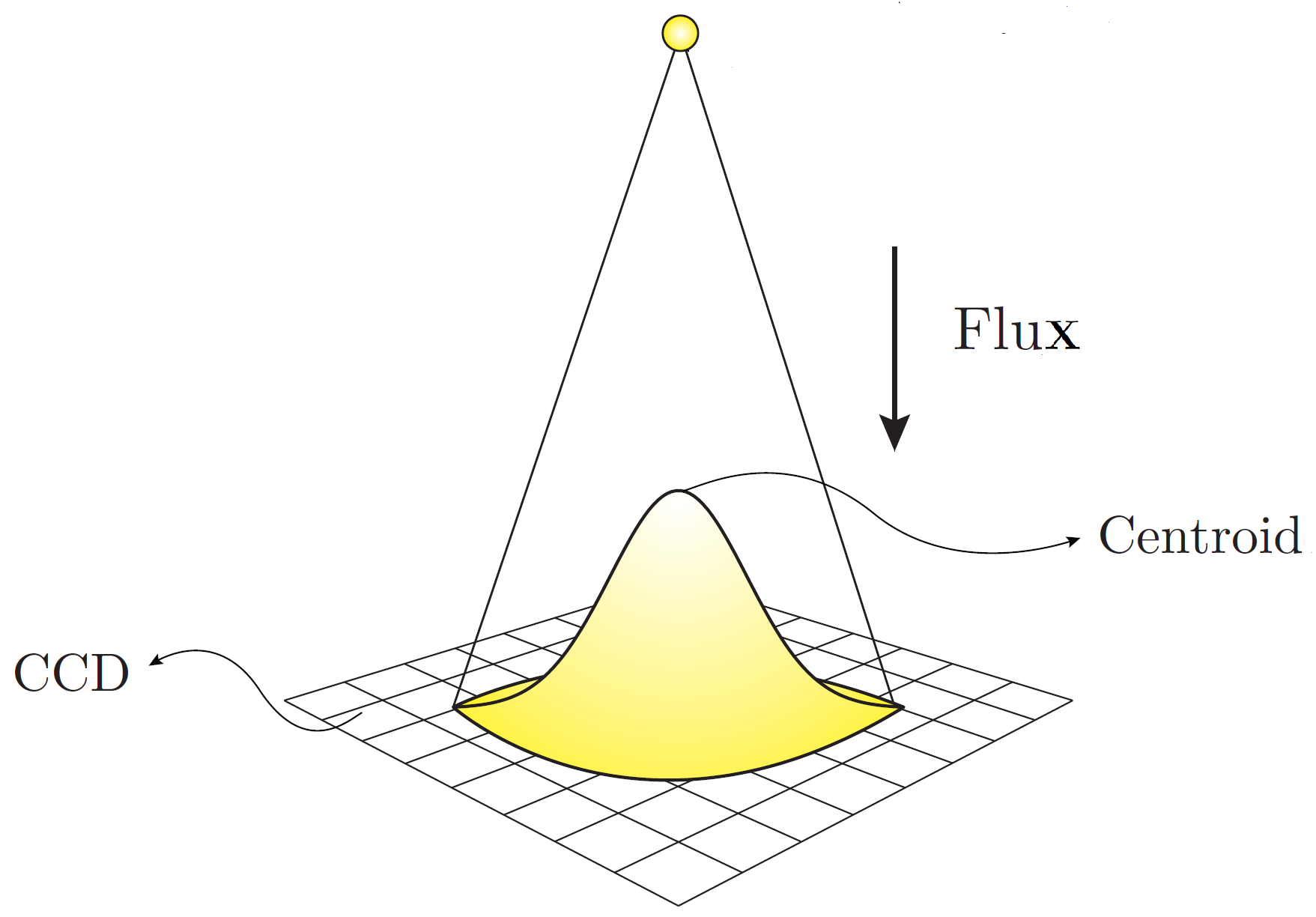}
    \caption{Illustration of a point source emitting flux onto a CCD array. The resulting intensity distribution on the detector follows the PSF, whose centroid corresponds to the position of the source.}
    \label{fig:centroid_formation}
\end{figure}

We now formalize the digital observational model introduced above. In the context of digital detectors, the observed data consist of discrete photon counts registered across a grid of pixels. Each measurement is affected by the underlying source intensity, the instrument's PSF, and additive background contributions.  The observed photon count at each pixel depends on the source flux, its position relative to the pixel grid, the PSF, and the background level. Figure~\ref{fig:centroid_formation} illustrates this process conceptually.

In the case of a point source, the object is described by two scalar parameters: its position $x_c \in \mathbb{R}$ within the detector array, and its total flux (brightness) $F \in \mathbb{R}^+$ (we focus here on a one-dimensional linear detector model; the extension to two-dimensional arrays is straightforward, involving a double integral over the pixel grid instead of a single one, as seen in analogous expressions such as Eq. (\ref{eq:pixel_response})). These parameters define a probability distribution $\mu_{x_c,F}$ over the observation space $\mathbb{X}$. The nominal intensity profile recorded on a photon-integrating device, such as a CCD, is then expressed as:
\begin{equation}
F_{x_c}(x) = F \cdot \phi(x - x_c, \sigma), \label{eq:profile_model}
\end{equation}
where $\phi(x - x_c, \sigma)$ is the one-dimensional normalized PSF. This nominal intensity profile reflects how the total flux $F$ is distributed over the detector according to the instrument’s PSF. The actual photon counts registered at each pixel are influenced by how this profile is sampled spatially, as well as by additional background contributions and the inherent randomness of photon arrival. In what follows, we examine in more detail the key components that define this observational model: the PSF, the total source flux, and the background noise.

\subsubsection{Source flux and noise contributions}

The nominal signal profile from Eq.~(\ref{eq:profile_model}) is not directly observable; it is perturbed by various noise sources that affect the recorded pixel values.
\begin{itemize}
    \item A physical background per pixel, generated by photon-related processes,
is given by
\begin{equation}
B_{\rm phys} = f_s \,\Delta x + D,
\label{eq:physical_background}
\end{equation}
where $f_s$ denotes the sky surface brightness and $D$ the dark current.
This corresponds to the mean number of photo-electrons accumulated in each
pixel, following standard CCD models
\citep{2001sccd.book.....J,2006hca..book.....H,2007ptd..book.....J,2008eiad.book.....M}.

In addition to the photon-generated background, the read-out noise ($RON$)
introduces a variance-only contribution. Although $RON^2$ is not a photon
flux, it is often folded into a single effective background term for
convenience in likelihood and Fisher-information formulations. Adopting
this convention, we define the effective background level as
\begin{equation}
B = B_{\rm phys} + RON^2,
\label{eq:effective_background}
\end{equation}
which serves as the Poisson-equivalent mean entering the noise model.
This formulation is compatible with commonly adopted CCD noise models in the
literature, where photon-generated terms and read-out noise are handled
jointly or separately depending on the chosen likelihood representation
(e.g., \citealt[eqs.~4--5]{prod2012impact}). 
We adopt this unified notation here solely for convenience in the subsequent
CRLB development.

    \item Photon noise, arising from the discrete and stochastic nature of photon arrivals, is modeled by treating the recorded photon count at each pixel as a Poisson random variable:
    \begin{equation}
    I_i \sim \text{Poisson}(\lambda_i(x_c, F)),
    \end{equation}
    where $\lambda_i(x_c, F)$ is the expected signal plus background at pixel $i$.

    \item Spatial sampling effects, due to the discrete nature of the detector grid, are introduced in the next subsection through the pixel integration process.
\end{itemize}

While the above formulation assumes that pixel noise is purely Poisson-distributed—a valid approximation in photon-limited regimes—real detectors introduce additional Gaussian readout noise. A more complete model thus corresponds to a Poisson–Gaussian mixture, leading to more intricate likelihood functions and estimation problems. Practical approaches such as the Anscombe transform~\citep{1948JRSS...Anscombe} and its generalizations to mixed Poisson–Gaussian noise~\citep{1995Murtagh} have been proposed to approximate this regime. For clarity and analytical tractability, we restrict our discussion here to the Poisson-limited case, deferring the mixed-noise treatment to future developments.

\subsubsection{Point spread function (PSF) and sampling}

A crucial factor in determining astrometric and photometric precision is the PSF, which describes the response of the imaging system to a point source. A commonly used model for the PSF is the Gaussian function:
\begin{equation}
\phi(x, \sigma) = \frac{1}{\sqrt{2\pi} \sigma} \exp\left(-\frac{x^2}{2\sigma^2}\right), \label{eq:gaussian_PSF}
\end{equation}
where $\sigma$ characterizes the spread of the light profile on the detector. This parameter depends on factors such as wavelength and atmospheric seeing~\citep{mendez2013,mendez2014}. This Gaussian approximation is appropriate for most ground-based observations dominated by atmospheric seeing, whereas space-borne instruments can approach diffraction-limited conditions where the PSF is no longer well described by a Gaussian and typically exhibits a complex structure with extended wings. A more detailed discussion on PSF modeling and fitting is provided in Subsection \ref{sec:psf_modelling}.
The detector samples the intensity profile through pixel integration. Let $\{x_i\}_{i \in \mathbb{Z}}$ be a uniform grid with spacing (pixel resolution) $\Delta x$. Then, the expected signal at pixel $i$ is given by:
\begin{equation}
\lambda_i(x_c,F) = F \cdot g_i(x_c) + B, \label{eq:expected_counts}
\end{equation}
where the pixel response function $g_i(x_c)$ integrates the PSF over pixel $i$:
\begin{equation}
g_i(x_c) = \int_{x_i - \Delta x/2}^{x_i + \Delta x/2} \phi(x - x_c, \sigma) \, dx. \label{eq:pixel_response}
\end{equation}
This pixel-integrated function is equivalent to the so-called effective PSF (ePSF) introduced by \citet{anderson2000toward}, which describes the combined effect of the optical PSF and the detector sampling. This simplified model neglects possible inter-pixel crosstalk and charge diffusion effects, which may modify the effective response in real detectors.

\begin{figure}[ht]
    \centering
    \includegraphics[width=0.7\textwidth]{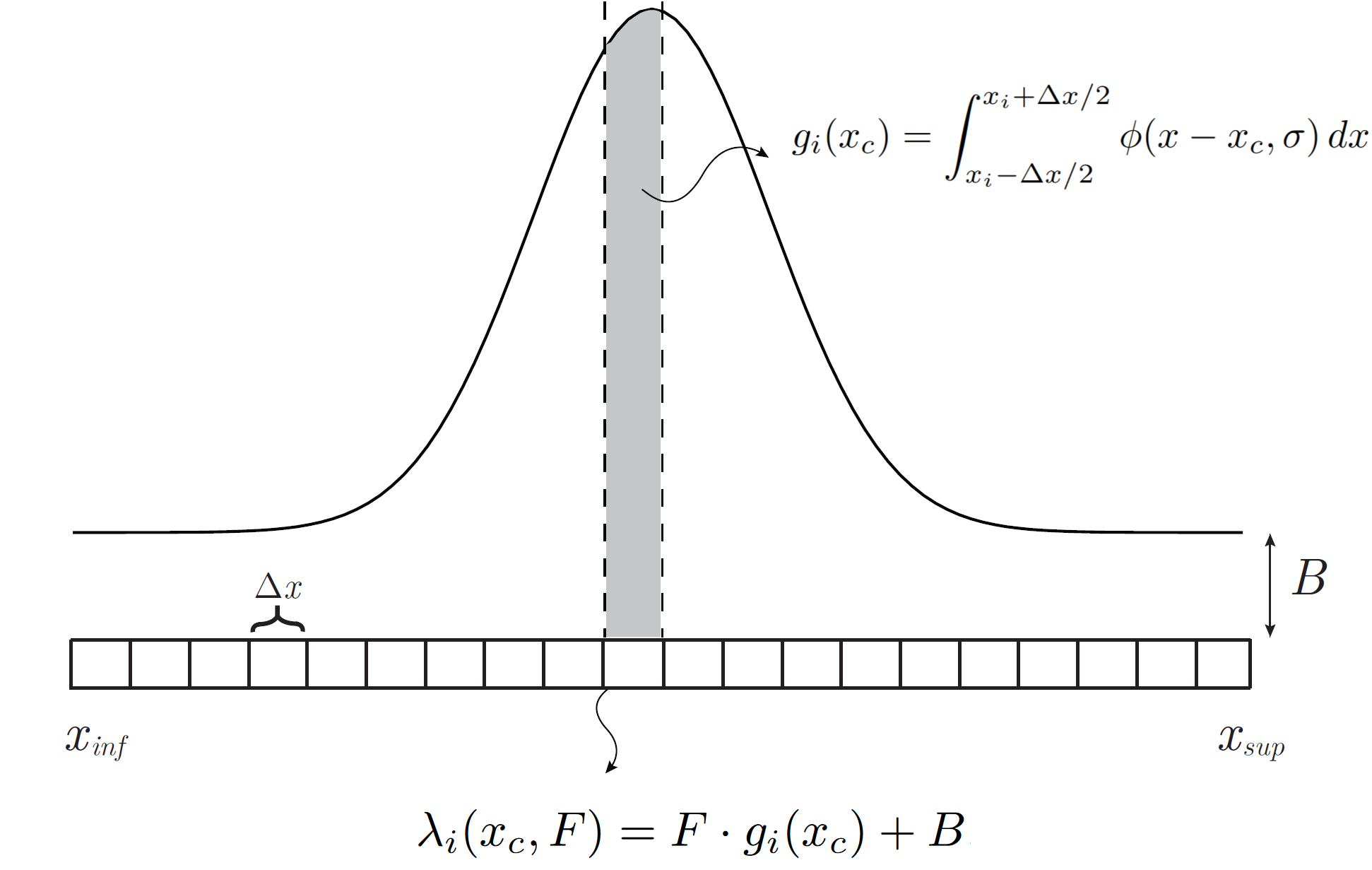}
    \caption{Pixel-integrated model of the observed intensity. The expected photon count at pixel $i$ is given by $\lambda_i(x_c, F) = F \cdot g_i(x_c) + B$, where $g_i(x_c)$ integrates the PSF over pixel $i$, and $B$ accounts for the background.}
    \label{fig:pixel_model}
\end{figure}

Figure~\ref{fig:pixel_model} illustrates how the continuous PSF is sampled by the detector grid, with each pixel integrating the light over its area to produce the expected photon count. A commonly used width measure for the PSF is the full-width at half-maximum (FWHM), related to $\sigma$ via:
\begin{equation}
FWHM = 2\sqrt{2\ln{2}} \, \sigma. \label{eq:FWHM}
\end{equation}
Together, the components described above define a complete observational model for a point source: each pixel produces a photon count that follows a Poisson distribution with mean $\lambda_i(x_c, F) = F \cdot g_i(x_c) + B$, where $g_i(x_c)$ captures the sampling of the PSF and $B$ models the background. Assuming independence between pixel measurements, the full dataset consists of a random vector $\mathbf{I} = \{I_1, \dots, I_n\}$ of independent observations. This probabilistic model serves as the foundation for statistical inference.

\subsubsection{The inference task}

Based on the observational model, the goal of inference is to recover the position $x_c$ and/or the total flux $F$ of a point source from the observed data $\mathbf{I}$. Assuming a finite set of $n$ pixels indexed by $\mathcal{N} = \{1, \dots, n\}$, centered around the source, the likelihood of the data under the parameter vector $(x_c, F)$ is given by:
\begin{equation}
L(\mathbf{I}; x_c, F) = \prod_{i =1}^n P_{\lambda_i(x_c,F)}(I_i), \label{eq:likelihood}
\end{equation}
where the Poisson probability mass function is:
\begin{equation}
P_{\lambda}(x) = \frac{e^{-\lambda} \cdot \lambda^x}{x!}. \label{eq:poisson_pmf}
\end{equation}
The task is then to construct an estimator $\tau(\cdot)$ that maps the vector of observations to an estimate of the source parameters:
\begin{equation}
\tau: \mathbb{N}^n \to \mathbb{R} \times \mathbb{R}^+, \qquad (\hat{x}_c, \hat{F}) \equiv \tau(\mathbf{I}).
\end{equation}
This estimation framework allows us to rigorously analyze the performance of different inference methods under varying observational conditions, which is the focus of the next sections. Although photon counts are discrete, the Poisson likelihood remains a continuous and differentiable function of the parameters $(x_c, F)$, fully compatible with the regularity conditions assumed in Fisher-information analysis.

\subsection{Practical implications for astrometric and photometric estimation}

The observational model developed above provides the foundation for precision analysis in both astrometry and photometry. In astrometry, the shape and width of the PSF, the background level, and the detector sampling resolution all influence the accuracy with which source positions can be estimated. In photometry, precise flux recovery depends on the ability to separate signal from background noise, and is affected by the estimation strategy and the choice of aperture size.

Understanding how these factors impact measurement precision enables the derivation of theoretical performance bounds, such as the CRLB, and informs the design of optimal estimation strategies. In the next sections, we explore the theoretical performance limits and evaluate the performance of widely used practical estimators in astrometric and photometric applications.

\subsection{Practical aspects of PSF modeling and recent advances}
\label{sec:psf_modelling}

While the Gaussian PSF assumption, introduced in Eq.~(\ref{eq:gaussian_PSF}),  provides a convenient analytical form to model and determne fundamental estimation principles, it is rarely fully adequate for the complexity required in modern astronomical settings. 
In practice, the PSF results from a combination of optical aberrations, diffraction, detector response, and—in the case of ground-based observatories—atmospheric turbulence. 
Because each of these factors imprints distinct spatial and chromatic signatures, the choice of PSF model has direct implications for the attainable astrometric and photometric precision. 
Over the years, PSF modeling has evolved through three complementary paradigms: (i) analytic formulations based on physical or phenomenological profiles, (ii) empirical pixel-integrated reconstructions, and (iii) data-driven or statistically regularized representations.

\subsubsection{Analytic representations}
Early modeling efforts adopted analytic functions that approximate the observed light distribution. 
The simplest form is the Gaussian profile, which is circularly symmetric and characterized by a width parameter that controls the concentration of light.
While the Gaussian offers mathematical simplicity and is directly linked to the diffusion of optical energy, it fails to reproduce the extended wings typically seen in real stellar images. A major advance in this direction was the introduction of the {Moffat profile~\citep{moffat1969theoretical},
\begin{equation}
\phi(r) \propto \left[1+\left(\frac{r}{\alpha}\right)^2\right]^{-\beta},
\end{equation}
where the exponent $\beta > 1$ regulates the wing amplitude: large $\beta$ yields a Gaussian-like profile, while smaller $\beta$ captures the turbulence-induced halo. 
Subsequent refinements, notably by \citet{trujillo2001effects}, quantified how atmospheric seeing modifies extended galaxy profiles convolved with Moffat PSF profiles. 

Hybrid models—mixtures of Gaussians, double Moffats, or power-law wings—were later adopted in survey pipelines (e.g. SDSS, DES, LSST) to describe the complex morphology of real PSFs across large fields~\citep{xin2018study}. 
These analytic families remain valuable because of their physical interpretability and closed-form normalization, yet they are insufficient once pixelization, undersampling, or detector inhomogeneities dominate the error budget.

\subsubsection{Empirical pixel-integrated reconstructions}
With the advent of space telescopes such as HST, the dominant challenge shifted from analytic fitting to accurate pixel-level representation. 
The effective PSF (ePSF) formalism introduced by \citet{anderson2000toward} for the HST/WFPC2 was transformative in this regard. 
Rather than modeling a continuous intensity distribution, the ePSF represents the detector-sampled PSF, incorporating the intra-pixel sensitivity and optical distortions through empirical reconstruction from high-SNR stellar images. 
This approach effectively models the probability that a photon is detected in a given pixel as a function of subpixel position, naturally accounting for undersampling and jitter effects. 
\citet{anderson2000toward} demonstrated that this formalism enabled sub-millipixel centroiding precision—an order-of-magnitude improvement over Gaussian or Moffat fits applied directly to sampled data.

The ePSF framework became the cornerstone of PSF calibration for high-precision missions. 
\citet{nardiello2022photometry} and \citet{libralato2024high} extended it to the JWST instruments (NIRCam and MIRI), introducing multidimensional grids to capture the dependence of the PSF on wavelength, field position, and temporal focus variations. Similarly, the Gaia consortium adopted a shapelet-based variant for its time-varying line-spread functions~\citep{rowell2021gaia}. 
These works collectively show that empirical, pixel-integrated models are indispensable when pursuing microarcsecond-level astrometry or sub-percent photometry, as they directly embed the instrumental response within the estimation framework.

\subsubsection{Data-driven representations}
In parallel, the availability of massive and heterogeneous imaging datasets has driven a transition toward statistical learning approaches that treat the PSF as a high-dimensional, smoothly varying function of instrumental and environmental parameters. \citet{nie2021point} introduced a Smooth Principal Component Analysis framework, in which PSFs are decomposed into orthogonal spatial modes $\phi_k(x,y)$ with coefficients that vary smoothly across the field through Laplacian regularization. Non-parametric extensions based on graph-Laplacian interpolation~\citep{schmitz2020euclid} follow the same spirit, ensuring that PSF variations remain consistent across irregular detector geometries and discontinuous CCD mosaics. In both cases, the key idea is to model the PSF as a continuous, spatially coherent field rather than as a collection of independent, localized profiles.

More recent efforts have integrated machine learning into PSF modeling and centroiding, effectively bypassing explicit analytic formulations. 
\citet{casetti2023star} demonstrated that convolutional neural networks trained on undersampled HST/WFPC2 images can directly infer stellar centroids from raw pixel intensities, removing the pixel-phase bias that affects classical ePSF methods, achieving sub-millipixel accuracy. 
At a broader scale, \citet{stone2023astrophot} introduced ASTROPHOT, a fully differentiable optimization framework that jointly fits PSF, background, and source parameters across entire images. 

\subsubsection{Current perspective}
Modern PSF modeling now spans a continuous range of approaches —from analytical and elegant solutions to empirical realism and, more recently, to adaptive, data-driven strategies. 

From the perspective of this work, such developments provide increasingly precise representations of the image-formation process, which serve as essential inputs to the statistical estimation of astrophysical parameters. 
Once the PSF is characterized—whether analytically, empirically, or through learned representations—the estimation of source position and flux remains governed by the same fundamental principles of statistical inference. 
The FIM and the CRLB continue to define the theoretical limits on attainable precision, and the results derived in the next section extend naturally to any differentiable PSF through its gradients with respect to the model parameters. 

The analysis presented in this review is intended as a theoretical framework for understanding and quantifying the ultimate precision achievable in astrometry and photometry. Although, for clarity and analytical simplicity, the numerical examples provided in later sections are based on a Gaussian PSF, it is important to stress that our analytical results are not restricted to this specific case. All key expressions, including those related to the FIM and the CRLB, are derived in terms of the pixel response function (see Eq.~\eqref{eq:pixel_response}), which depends on the derivatives of the PSF with respect to the model parameters. As such, the framework we develop is fully general and directly applicable to any differentiable PSF. In the following section, we elaborate on these foundations by introducing the FIM and the CRLB as the fundamental statistical tools that set the precision limits for astrometric and photometric estimators.


\section{Fundamental bounds in astrometry and photometry} \label{sec:bounds} 

\subsection{Early contributions in precision limits}

Early investigations into the precision limits of astrometric measurements can be traced back to the seminal works by \citet{vanaltena1975} and \citet{lindegren1978}, who explored how factors such as the PSF, detector sampling, and background noise affect positional accuracy.

The work of \citet{king1983accuracy} marked a meaningful step in conceptual understanding, who formulated a quantitative framework for estimating the precision of brightness and position measurements of stellar images recorded on digital detector arrays. King developed analytical expressions for the expected variance in photometry and astrometry under Poisson statistics, introducing the notion of an equivalent noise area—a practical way to assess the contribution of background noise as a function of the PSF morphology. The study also rigorously analyzed how finite pixel size and image centering relative to the pixel grid impact measurement accuracy, providing tabulated corrections for various observational regimes. 

Building on these foundations, \citet{winick1986} derived analytical expressions for the CRLB in the context of position estimation using CCD detectors. His work quantified how the trade-off between pixel size and PSF width influences astrometric precision, identifying optimal configurations that minimize the estimator variance. Around the same time, \citet{chen1987} extended this analysis to the joint estimation of position and velocity in CCD-based optical tracking systems, deriving CRLBs under realistic sensor and noise models. These results underscored the applicability of CRLB-based analysis to the design and evaluation of high-precision optical instruments.

In parallel, \citet{jakobsen1992cramer} applied the CRLB framework to the problem of two-dimensional stellar photometry using aberrated Hubble Space Telescope images. They derived a simple yet general expression for the minimum achievable photometric error, valid for arbitrary PSFs and degrees of crowding, independent of the deconvolution or processing techniques applied. This analysis demonstrated how photon statistics impose a fundamental limit on precision, and it further quantified the irrecoverable sensitivity loss introduced by the spherical aberration of the Hubble Space Telescope's Faint Object Camera.

A significant step forward came with the work of Hans-Martin~\citet{adorf1996limits}, who systematically introduced the use of the CRLB in observational astrophysics. His contributions emphasized the central role of Fisher information in astrometry, photometry, and spectral analysis, and positioned the CRLB as a practical tool for assessing the efficiency of estimators. He also highlighted the relevance of the CRLB beyond estimation theory, showing its applicability to instrument design, where it can guide choices by quantifying their impact on achievable precision.

In modern formulations, the CRLB establishes a fundamental lower limit on the variance of any unbiased estimator of a given parameter. As such, it provides a rigorous benchmark for evaluating the best possible performance achievable under a given observational model.

Here, we present a unified treatment of the CRLB in the context of astrometry and photometry, following the formulations developed by \citet{mendez2013}, \citet{espinosa2018optimality}, and \citet{vicuña}. We consider both scalar and joint estimation settings, where the parameter of interest may be one-dimensional, such as position in astrometry, or multidimensional, involving simultaneous estimation of multiple quantities such as position and flux, or flux and background. A unified notation is introduced to facilitate direct comparison across these different inference problems and observational regimes.

\subsection{General formulation of the Cramér-Rao bound}

The notorious CRLB offers a performance bound on the variance estimation error of the family of unbiased estimators\footnote{In the sense that $\mathbb{E}\{ \hat{\boldsymbol{\theta}} \} = \boldsymbol{\theta}$.}. We revisit the multidimensional\footnote{In the sense that various parameters are being determined.} version of this result:

\begin{teo} \citep{rao1945, cramer1946} 
Let $\mathcal{N}$ a finite set of integers with $|\mathcal{N}|=n$, $\mathbf{I}=\{I_{i}\}_{i \in \mathcal{N}}$ be a collection of observations, whose likelihood function $L(\cdot; \boldsymbol{\theta})$ is induced by a parameter vector $\boldsymbol{\theta} = (\theta_{1},\dots,\theta_{m}) \in \boldsymbol{\Theta}, m \in \mathbb{N}$ over a parameter space $\boldsymbol{\Theta}$ (typically $\boldsymbol{\Theta} = \mathbb{R}^{m}$), such that the following ``regularity condition" is satisfied:
\begin{equation}
\mathbb{E}\left\{ \frac{\partial \ln{L(\mathbf{I};\boldsymbol{\theta})}}{\partial \theta_{i}} \right\} = 0 \,, \hspace{1mm} \forall i \in \{1,\dots,m\} \,, \forall \boldsymbol{\theta} \in \boldsymbol{\Theta}. \label{eq:CRLB_reg}
\end{equation}

(i) Then, any unbiased estimator $\hat{\boldsymbol{\theta}}$ of $\boldsymbol{\theta}$, given by a regression rule $\tau: \mathbb{N}^{n} \to \boldsymbol{\Theta}, \hat{\boldsymbol{\theta}} \equiv \tau(\mathbf{I})$, has a covariance matrix $K_{\hat{\boldsymbol{\theta}}}\equiv\mathbb{E} \left\{ \left( \hat{\boldsymbol{\theta}} - \boldsymbol{\theta}\right) \cdot \left( \hat{\boldsymbol{\theta}} - \boldsymbol{\theta} \right)^{\dagger} \right\}$ that satisfies\footnote{$\succeq \mathbf{0}$ means that the matrix on the left-hand side of Eq.~(\ref{eq:CRLB_psd}) is positive semi-definite: $\forall x \in \mathbb{R}^m$, $x^\dagger \cdot (K_{\hat{\boldsymbol{\theta}}} - \mathcal{I}_{\boldsymbol{\theta}}^{-1})\cdot x \geq 0$.}:
\begin{equation} \label{eq:CRLB_psd}
K_{\hat{\boldsymbol{\theta}}} - \mathcal{I}_{\boldsymbol{\theta}}^{-1} \succeq \mathbf{0}.
\end{equation}
$\mathcal{I}_{\boldsymbol{\theta}} \in \mathcal{M}^{m \times m}$ denotes the FIM, whose components are defined by: $\forall i,j \in \{1,\dots,m \},$
\begin{equation}
[\mathcal{I}_{\boldsymbol{\theta}}]_{(i,j)} 
\equiv -\mathbb{E} \left\{ \frac{\partial^{2} \ln{L(\mathbf{I};\boldsymbol{\theta})}}{\partial \theta_{i} \partial \theta_{j}} \right\} \,.
\label{eq:fisher_matrix}
\end{equation}

(ii) Furthermore, if there exists a function $\mathbf{h}: \mathbb{R}^n \to \boldsymbol{\Theta}$ such that: 
\begin{equation}
\frac{\partial \ln{L(\mathbf{I};\boldsymbol{\theta})}}{\partial \theta_{i}} = \left[ \mathcal{I}_{\boldsymbol{\theta}} (\mathbf{h}(\mathbf{I}) - \boldsymbol{\theta}) \right]_{i} \,, \hspace{3mm} \forall i \in \{1,\dots,m\} \,, \label{eq:CRLB_g_cond}
\end{equation}
then the minimal variance unbiased estimator is given by $\hat{\boldsymbol{\theta}} = \mathbf{h} (\mathbf{I})$, and its optimal covariance matrix is $\mathcal{I}_{\boldsymbol{\theta}}^{-1}$.
\label{theo:CRLB}

\end{teo}
From Theorem \ref{theo:CRLB}, any unbiased estimator $\hat{\theta}_i$ of $\theta_i$ satisfies that (from Eq.~(\ref{eq:CRLB_psd})):
\begin{equation}
Var[\hat{\theta}_{i}] = [K_{\hat{\boldsymbol{\theta}}}]_{(i,i)} \geq [\mathcal{I}_{\boldsymbol{\theta}}^{-1}]_{(i,i)} \,, \hspace{1mm} \forall i \in \{1,\dots,m\}, 
\end{equation}
therefore $[\mathcal{I}_{\boldsymbol{\theta}}^{-1}]_{(i,i)}$ is a lower bound for the mean squared error (MSE) of $\hat{\theta}_i$.\footnote{Theorem~1 does not guarantee the existence of an estimator that achieves the CRLB, however, it is still possible to find the minimum variance unbiased estimator \citep{kay1993fundamentals}.}

\subsection{CRLB for the estimation of the position -- one-dimensional case}

We begin with the classical setting where only the position \( x_c \) of a point source is unknown, while its flux \( F \) and the background \( B \) are assumed known. The precision limit for estimating \( x_c \) requires computing the Fisher information of the position parameter under a Poisson noise model.

Following the formulation of~\citet[Proposition 3]{mendez2013}, the Fisher information for estimating the source position is
\begin{equation}
\mathcal{I}_{x_c} = \sum_{i =1}^n \frac{F^2 \left( \frac{d g_i(x_c)}{dx_c} \right)^2}{F g_i(x_c) + B}.
\end{equation}
Therefore, the CRLB for any unbiased estimator \( \hat{x}_c \) is given by
\begin{equation}
Var[\hat{x}_c] \geq \sigma^2_{x_c}\equiv \frac{1}{\mathcal{I}_{x_c}}.
\label{crbound}
\end{equation}
This expression reveals that the positional accuracy is fundamentally limited by the source brightness \( F \), the background \( B \), and the gradient of the PSF with respect to \( x_c \). This bound has become a reference for assessing astrometric performance under realistic observational conditions.

The work by \citet{mendez2013} provides a systematic and in-depth analysis of this bound in the one-dimensional case. Beyond deriving the Fisher information under a Poisson noise model, the paper thoroughly investigates how the CRLB depends on key system parameters such as the pixel size, SNR, PSF, and the subpixel location of the source. Closed-form approximations are derived for both background-limited and source-dominated regimes, revealing that the astrometric uncertainty scales as $ B/F^2 $ in the former and $1/F $ in the latter. Their study highlights the role of undersampling and source decentering, showing that dithering can mitigate resolution losses. 

Moreover, theoretical predictions are compared with results from ground- and space-based observatories, demonstrating that practical estimators such as maximum likelihood can closely approach the CRLB under a wide range of observational conditions. Then, their work has become an essential reference for the design and evaluation of high-precision astrometric measurements settings.

In addition, a particularly illustrative behavior emerges when analyzing the CRLB as a function of the pixel size $\Delta x$. As first noted by \citet{mendez2013}, the bound exhibits a characteristic “valley” or U-shape: for very small pixels, positional precision deteriorates not because the total background flux increases—which depends mainly on the detector area—but because subdividing the same region distributes the signal over a larger number of pixels, each contributing its own background and instrumental noise term. Conversely, for very large pixels, resolution is lost due to coarse sampling of the PSF. The existence of an optimum region in $\Delta x$ reflects a fundamental trade-off between cumulative noise and spatial resolution, which directly guides detector design. In practice, this optimum is relatively broad, meaning that small deviations in sampling within this region have limited impact on the achievable astrometric precision.

This effect is further nuanced by the fact that the dominant background often originates from diffuse sky brightness rather than detector noise, implying that the background per pixel scales with $\Delta x$. This contrasts with other applications (e.g., fast-readout sensors in military or daytime contexts), where the pixel background may remain independent of pixel size. For astronomical CCDs, accounting for this scaling is essential: the optimum pixel size must balance the competing requirements of maximizing signal-to-noise ratio and retaining sufficient sampling of the PSF. Figure~\ref{fig:crlb_pixel} illustrates this trade-off, showing how the CRLB defines a practical design window for pixel sizes that minimize astrometric uncertainty.

\begin{figure}[ht!]
    \centering
    \includegraphics[width=0.8\textwidth]{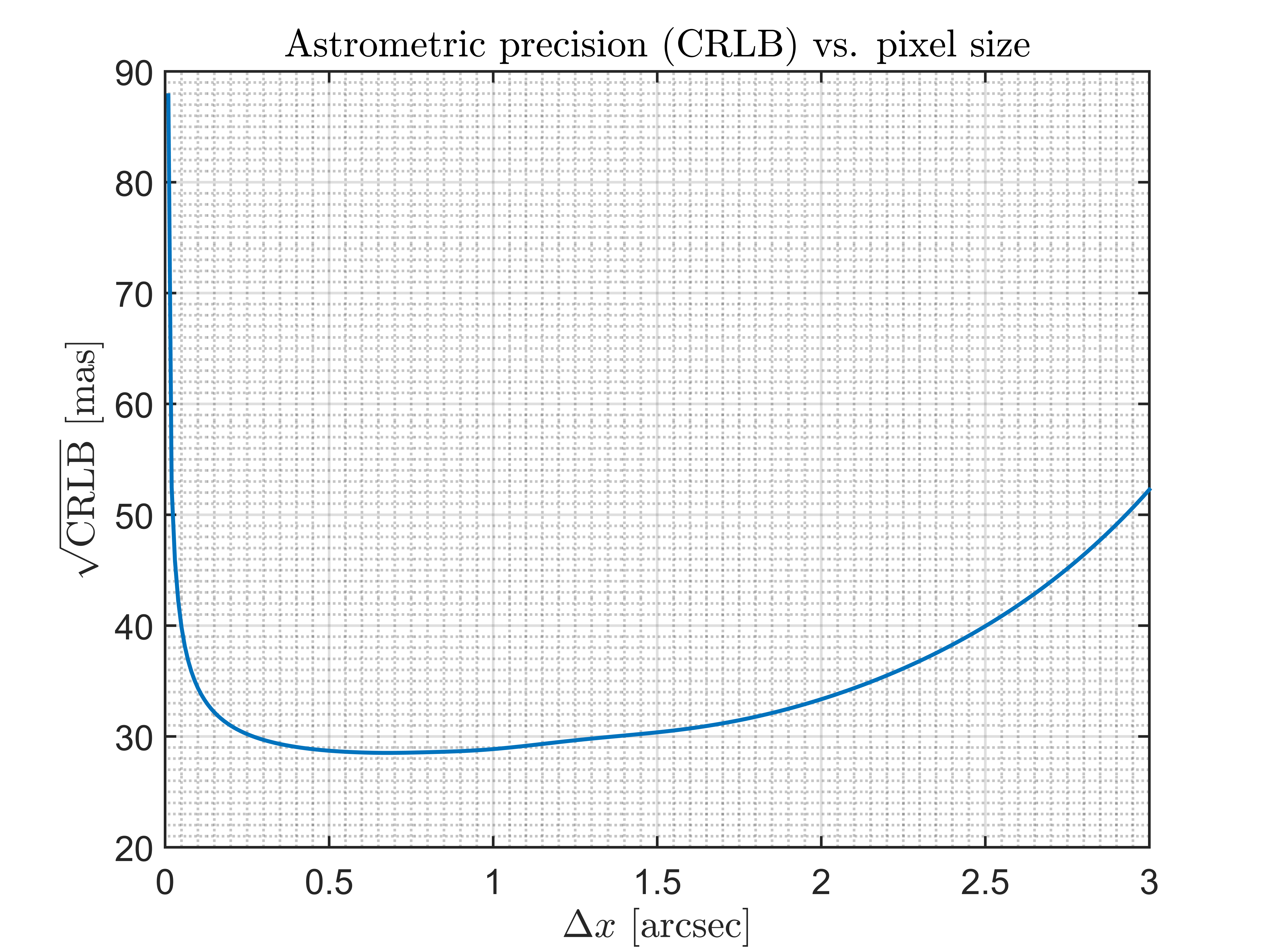}
    \caption{Square root of the CRLB for centroid estimation as a function of the pixel size $\Delta x$. Parameters correspond to a typical ground-based observation: total source flux $F=1000\ e^-$, background level $B=10\ e^-$ per pixel, Gaussian PSF with standard deviation $\sigma = 0.8$ arcsec, centroid $x_c=1.12$ pixels, and a well-sampled window of $n=201$ pixels. The CRLB is expressed in milliarcseconds (mas). The curve shows the characteristic U-shape, with an optimum region in $\Delta x$ that minimizes the astrometric uncertainty.}
    \label{fig:crlb_pixel}
\end{figure}

\subsection{CRLB for the joint estimation of position and source flux}

The previous analysis focused on the classical one-dimensional setting in which only the source position $x_c$ is unknown, with flux $F$ and background $B$ considered fixed and known. We now consider more general and realistic case where both $x_c$ and $F$ must be estimated simultaneously. This requires computing the full FIM associated with the parameter vector $\boldsymbol{\theta} = (x_c, F)$, which accounts for the statistical coupling between astrometric and photometric uncertainties. As detailed in~\citet[Proposition 3]{mendez2014}, the FIM in this case is given by the following expression:
\begin{equation}
\mathcal{I}_{\boldsymbol{\theta}} =
\begin{bmatrix}
\sum\limits_{i =1}^n \frac{\left( \frac{\partial \lambda_i(x_c,F)}{\partial x_c} \right)^2}{\lambda_i(x_c,F)} &
\sum_{i =1}^n \limits\frac{\frac{\partial \lambda_i(x_c,F)}{\partial x_c} \cdot g_i(x_c)}{\lambda_i(x_c,F)} \\
\sum_{i =1}^n \limits\frac{\frac{\partial \lambda_i(x_c,F)}{\partial x_c} \cdot  g_i(x_c) }{\lambda_i(x_c,F)} &
\sum_{i =1}^n \limits\frac{ g_i^2(x_c)}{\lambda_i(x_c,F)}
\end{bmatrix}.
\end{equation}
The non-zero off-diagonal terms in the Fisher matrix account for the statistical coupling between the estimation of $x_c$ and $F$, reflecting how uncertainty in one parameter propagates to the other. These expressions generalize the CRLB for position-only estimation and allow one to evaluate astrometric accuracy when the source flux is also unknown.

\subsection{CRLB for the joint estimation of source flux and background} \label{sec:CRLB_analysis}

A recent contribution from \citet{vicuña} presents a detailed analysis of the CRLB for the joint estimation of source flux and background level, under the assumption that the astrometric position $x_c$ is known. We summarize their main result below:

\begin{lema}\citep[Lemma 3.1]{vicuña}\label{lm_fisher_joint}
If the astrometry $x_c \in \mathbb{R}$ is fixed and known, and we want to estimate the pair $(F, B)$ from $\mathbf{I} \sim   L(\mathbf{I} ;F, B)  $ in (\ref{eq:likelihood}), then the FIM in Eq.~(\ref{eq:CRLB_psd}) is given by:
\begin{equation}
\mathcal{I}_{\boldsymbol{\theta}} = 
\begin{bmatrix}
\mathcal{I}_{1,1} & \mathcal{I}_{1,2} \\
\mathcal{I}_{2,1} & \mathcal{I}_{2,2}
\end{bmatrix}
\equiv 
\begin{bmatrix}
\sum_{i =1}^n \limits\frac{g_{i}^{2}(x_{c})}{\lambda_{i}(x_c,F)} & 
\sum_{i =1}^n \limits\frac{g_{i}(x_{c})}{\lambda_{i}(x_c,F)} \\
\sum_{i =1}^n \limits\frac{g_{i}(x_{c})}{\lambda_{i}(x_c,F)} & 
\sum_{i =1}^n \limits\frac{1}{\lambda_{i}(x_c,F)}
\end{bmatrix} \,. 
\label{eq:FB_fisher}
\end{equation} 
\end{lema}
Consequently, from Theorem \ref{theo:CRLB}, the following two-dimensional CRLB for any pair of unbiased estimates $\hat{F}$ and $\hat{B}$ can be respectively defined as:
\begin{equation}
Var[\hat{F}]  \geq \sigma_{F}^{2}  = \frac{\mathcal{I}_{2,2}}{\mathcal{I}_{1,1} \cdot \mathcal{I}_{2,2} - \mathcal{I}_{1,2}^{2}} = \frac{\mathcal{I}_{2,2}}{|\mathcal{I}_{\boldsymbol{\theta}}|} \,,\label{eq:FBound}
\end{equation}
\begin{equation}
Var[\hat{B}] \geq \sigma_{B}^{2} = \frac{\mathcal{I}_{1,1}}{\mathcal{I}_{1,1} \cdot \mathcal{I}_{2,2} - \mathcal{I}_{1,2}^{2}} = \frac{\mathcal{I}_{1,1}}{|\mathcal{I}_{\boldsymbol{\theta}}|} \,. \label{eq:BBound}
\end{equation}
It is worth noting that the statistical interaction (correlation) between the unknown components of the problem is captured
by the off-diagonal elements of the FIM, i.e., $\mathcal{I}_{1,2}$ and $\mathcal{I}_{2,1}$, in Eqs.~(\ref{eq:FBound}) and~(\ref{eq:BBound}). Importantly, in the special case that $\mathcal{I}_{1,2} = 0$, the estimates become decoupled and one could consider that the joint estimation task reduces to two isolated 1D estimation problems. Related analyses were presented by \citet{gai2017performance}, who investigated the simultaneous estimation of flux, background, and photocenter in one-dimensional models, discussing its practical implications for instrument calibration and precision assessment.

\subsection{Estimators and their proximity to the CRLB}
The CRLB provides a fundamental benchmark for evaluating the performance of any unbiased estimator. It defines a lower limit on the variance that any such estimator can achieve, given a statistical model. In general, for non-linear or Poisson-based models, such as those encountered in photometry and astrometry, no unbiased estimator exists that exactly achieves the CRLB \citep{mendez2013,lobos2015}. However, asymptotically efficient estimators, such as the Maximum Likelihood (ML) Estimator, can approach the CRLB under regularity conditions and in the high-SNR regime as studied by \citet{kay1993fundamentals} and \citet{espinosa2018optimality}.

In astronomical practice, the Least-Squares (LS) estimator \citep{kay1993fundamentals} has historically been widely used due to its computational simplicity and intuitive appeal. Nonetheless, LS is known to be suboptimal under heteroscedastic or non-Gaussian noise conditions \citep{lobos2015}. In contrast, ML estimation is asymptotically efficient and better suited for Poisson-dominated settings, where noise scales with the signal \citep{espinosa2018optimality}. Addresing this comparison, a detailed comparison between LS and ML estimators in astrometric applications was carried out by~\citet{mendez2013}, revealing significant differences in estimator performance, especially for low SNRs and asymmetric PSFs.

In fact, the transition from LS to ML-based approaches has been pivotal in modern astrometric missions. A paradigmatic example is the Gaia mission, where the core astrometric solution abandoned traditional LS in favor of a global ML estimation framework to better account for complex noise and calibration models~\citep{lindegren2008}. This shift was critical for achieving microarcsecond-level accuracy across the sky.

Moreover, recent studies have shown that joint inference strategies and weighted LS methods, where the weights are tuned according to the local noise properties, can produce variances close to CRLB even in challenging observational scenarios~\citep{mendez2014}. These methods offer a practical compromise when full ML estimator implementation is computationally prohibitive.

Although the CRLB may not always be attainable by an unbiased estimator, it serves as a standard for performance evaluation. In practice, the choice of estimator (LS, weighted LS, or ML) impacts how closely one can approach this limit. The next sections delve into the empirical behavior of these commonly used estimators in astrometry and photometry, examining how closely their performance aligns with the CRLB under different observational conditions.

\section{Performance analysis and estimators in astrometry} \label{sec:astro_estimators} 

In this section, we focus on the empirical performance of classical estimators in astrometry, examining how they compare against theoretical precision bounds. Emphasis is placed on the LS and ML estimation methods, which are commonly used in practice. We discuss their relative strengths and limitations under typical observational conditions, setting the stage for understanding the extent to which they approach the CRLB in realistic scenarios.

\subsection{Suboptimality of the least squares estimator}

\citet{lobos2015} analyze the performance of the LS estimator for estimating the position of a point source. The LS estimator is defined as the minimizer of the squared difference between the observed photon counts and their expected values under the model. More precisely, let $\mathbf{I} = \{I_1, \dots, I_n\}$ denote the observed photon counts and let us consider the cost function:
\begin{equation}
J_{ls}(\mathbf{I}, \alpha) = \sum_{i=1}^{n} (\lambda_i(\alpha,F) - I_i)^2.
\end{equation}
Then, the LS estimator for the position parameter $x_c$ is given by the solution of:
\begin{equation}
\tau_{\text{LS}}(\mathbf{I}) = \arg\min_{\alpha \in \mathbb{R}} J_{ls}(\mathbf{I}, \alpha).
\label{impli}
\end{equation}
Their main result by \cite{lobos2015} is presented in the following theorem:
\begin{teo}\citep[Theorem 1]{lobos2015}
\label{lobost}
Let $x_c \in \mathbb{R}$ be a fixed but unknown parameter, and let $\mathbf{I} \sim L(\mathbf{I}; x_c)$ in (\ref{eq:likelihood}). 
Under regularity conditions on the LS cost function (see \citealt{lobos2015} for the detailed assumptions), the bias and variance of the LS estimator $\tau_{\mathrm{LS}}(\mathbf{I})$ satisfy
\begin{equation}
\left| \mathbb{E}\{\tau_{\mathrm{LS}}(\mathbf{I})\} - x_c \right| \leq \epsilon(\delta),
\label{boundone}
\end{equation}
\begin{equation}
\mathbb{E}\left\{ (\tau_{\mathrm{LS}}(\mathbf{I}) - x_c)^2 \right\} \in 
\left[ 
\frac{\sigma^2_{\mathrm{LS}}(n)}{(1+\delta)^2}, \,
\frac{\sigma^2_{\mathrm{LS}}(n)}{(1-\delta)^2}
\right],
\label{boundtwo}
\end{equation}
where
\begin{equation}
\sigma^2_{\mathrm{LS}}(n) \equiv 
\frac{\mathbb{E}\{J'_{\mathrm{LS}}(\mathbf{I}, x_c)^2\}}
{(\mathbb{E}\{J''_{\mathrm{LS}}(\mathbf{I}, x_c)\})^2},
\end{equation}
and $\delta \in (0,1)$.
\end{teo}
Importantly, Theorem \ref{lobost} provides analytical bounds in Eq. (\ref{boundone}) and (\ref{boundtwo}) for both the bias and the MSE of the LS estimator, offering a refined understanding of its performance across different SNR regimes. One of the key insights derived from the closed form expressions of this theorem is that the LS estimator exhibits varying efficiency depending on the SNR conditions of the problem.

They show that the LS estimator is nearly optimal in the low-SNR regime, as its performance closely aligns with the CRLB. This result suggests that, under conditions where the observed signal is weak relative to noise, the LS method provides an accuracy nearly as good as the best unbiased estimator possible. The near-optimality in this regime reinforces the continued use of LS techniques in cases where noise dominates the measurement process, making it a reliable choice for astrometric position estimation in challenging observational conditions.

Conversely, the evaluation of Theorem \ref{lobost} show that LS estimator becomes significantly suboptimal at high SNRs compared to the CRLB (this is clearly shown empirically, e.g., in Figure~14 of \citet{bouquillon2017characterizing}). This inefficiency from evaluating the expression in Theorem \ref{lobost} can be atributed to the fact that the LS method does not account for the statistical properties of the Poisson noise present in astrometric measurements in high SNR. As a result, the estimator exhibits a higher MSE than the theoretical lower bound, indicating that alternative estimation approaches—such as maximum likelihood estimation or weighted LS methods—may be more suitable for high-SNR scenarios.

\subsection{Optimality of the maximum likelihood estimator}

\citet{espinosa2018optimality} establishes general bias and variance bounds for estimators that are defined as solutions to optimization problems. This framework is particularly useful when the estimator solution of a problem like Eq. (\ref{impli}) has no closed-form expression, which is often the case in astrometry. Remarkably, this richer result applies to both Weighted Least Squares (WLS) and ML estimators, allowing for an analytical comparison of their performance. The general result is presented below.
\begin{teo}\citep[Theorem 1]{espinosa2018optimality}
\label{espit}
Let $x_c \in \mathbb{R}$ be an unknown parameter, and let $\mathbf{I}\sim L(\mathbf{I}; x_c)$ in (\ref{eq:likelihood}). 
For a broad class of estimators $\tau_J(\mathbf{I})$ defined implicitly as the minimizers of differentiable cost functions $J(\alpha, \mathbf{I})$, there exist positive constants $\epsilon_J(n)$ and $\beta_J(n)$ such that
\begin{equation}
\left| \mathbb{E}\{\tau_J(\mathbf{I})\} - x_c \right| \leq \epsilon_J(n),
\end{equation}
\begin{equation}
Var[\tau_J(\mathbf{I})] \in \left( \sigma_J^2(n) - \beta_J(n),\ \sigma_J^2(n) + \beta_J(n) \right).
\end{equation}
where $\sigma_J^2(n)$ represents the nominal variance associated with $J$.
The required technical assumptions can be found in \citet{espinosa2018optimality}.
\end{teo}
This result provides general conditions under which the performance of an estimator can be analytically bounded in terms of bias and variance. In particular, it guarantees that any estimator satisfying the stated regularity conditions will exhibit a bias bounded by $\epsilon_J(n)$ and a variance that stays within a small neighborhood, defined by $\beta_J(n)$, of a nominal variance $\sigma_J^2(n)$. Crucially, this framework unifies the treatment of different estimators, including both WLS and ML, allowing for a systematic comparison under a common analytical structure. In what follows, we apply these results to the WLS and the ML estimators to show their relevance in astrometry.

Let $\mathbf{I} = \{I_1, \dots, I_n \}$ denote the observed photon counts and consider the weighted cost function:
\begin{equation}
J_w(\mathbf{I}, \alpha) = \sum_{i=1}^{n} w_i \cdot \left( \lambda_i(\alpha, F) - I_i \right)^2,
\label{wlss}
\end{equation}
where $w_1, \dots, w_n$ are non-negative weights that reflect the relative importance or reliability of each pixel measurement.
Then, the WLS estimator for the position parameter $x_c$ is defined as:
\begin{equation}
\tau_{\text{WLS}}(\mathbf{I}) = \arg\min_{\alpha \in \mathbb{R}} J_w(\mathbf{I}, \alpha).
\label{costwls}
\end{equation}
The corresponding performance of the WLS estimator within this framework is summarized in the following result:

\begin{teo}\citep[Theorem 2]{espinosa2018optimality}
\label{espitdW}
Let $\tau_{\text{WLS}}(\mathbf{I})$ be the WLS defined above in Eq. (\ref{costwls}). Then,
\begin{equation}
\left| \mathbb{E}\{\tau_{\text{WLS}}(\mathbf{I})\} - x_c \right| \leq \epsilon_{\text{WLS}}(n),
\end{equation}
\begin{equation}
Var[\tau_{\text{WLS}}(\mathbf{I})] \in \left( \sigma_{\text{WLS}}^2(n) - \beta_{\text{WLS}}(n),\ \sigma_{\text{WLS}}^2(n) + \beta_{\text{WLS}}(n) \right),
\end{equation}
where the nominal variance $\sigma_{\text{WLS}}^2(n)$ is:
\begin{equation}
\sigma_{\text{WLS}}^2(n) =   \frac{\sum_{i=1}^n w_i^2\lambda_i(x_c,F) (\frac{d \lambda_i(x_c,F)}{d x_c})^2}{\left ( \sum_{i=1}^n w_i^2\left (\frac{d \lambda_i(x_c,F)}{d x_c}\right )^2 \right )^2} .
\label{nominalvarwls}
\end{equation}
The explicit expressions for $\epsilon_{\text{WLS}}(n)$ and $\beta_{\text{WLS}}(n)$ can be derived from the general bounds given in \citet{espinosa2018optimality}, but they involve lengthy combinations of higher-order moments and derivatives of $\lambda_i(x_c, F)$ which are omitted here for brevity.
\end{teo}
If we look at the nominal variance of the WLS estimator derived in Theorem \ref{espitdW}, as expressed in Eq.~(\ref{nominalvarwls}), and compare it to the CRLB in Eq.~(\ref{crbound}), we observe that their structures are remarkably similar.  The WLS variance equals the CRLB if and only if the weights are chosen as $w_i = K / \lambda_i(x_c, F)$ for some constant $K > 0$. This optimal weighting assignment, however, depends on the true (and unknown) source position $x_c$, making it fundamentally unfeasible in practice. As such, any attempt to select weights without prior knowledge of $x_c$ will inevitably result in suboptimal performance relative to the CRLB. For clarity, the weights $w_i$ in Eq.~(\ref{wlss}) are fixed a priori and do not depend on the estimated parameters; the adaptive case is discussed separately in Section~\ref{awlssec}.


This dependency of the weighting scheme on the unknown source parameters highlights an intrinsic limitation of the WLS estimator: it cannot uniformly achieve the CRLB across all possible source positions, particularly in high-SNR regimes. A similar phenomenon was observed by \citet{lobos2015} for the unweighted LS estimator, whose variance diverges significantly from the CR bound as the SNR increases. This comparison illustrates the broader tendency of least-squares estimators to lose efficiency when the assumed noise model or weighting does not perfectly match the underlying statistics. 
These considerations motivate the use of the ML estimator, which, under regularity conditions, is known to achieve the CR bound asymptotically.

The ML estimator is defined as the value that maximizes the likelihood function under the assumed statistical model. More precisely, for the Poisson distribution described earlier, the ML estimator for the position $x_c$ is obtained by minimizing the negative log-likelihood:
\begin{equation}
\tau_{\text{ML}}(\mathbf{I}) = \arg \min_{\alpha \in \mathbb{R}} \underbrace{\sum_{i=1}^{n} -I_i \ln(\lambda_i(\alpha, F)) + \lambda_i(\alpha, F)}_{J_{ML}(\mathbf{I},\alpha)}.
\label{mlcost}
\end{equation}
When Theorem \ref{espit} is applied to the ML estimator, we obtain the following:

\begin{teo}\citep[Theorem 3]{espinosa2018optimality}
\label{espitd}
Let $\tau_{\text{ML}}(\mathbf{I})$ be the ML estimator in Eq. (\ref{mlcost}). Then,
\begin{equation}
\left| \mathbb{E}\{\tau_{\text{ML}}(\mathbf{I})\} - x_c \right| \leq \epsilon_{\text{ML}}(n),
\end{equation}
\begin{equation}
Var[\tau_{\text{ML}}(\mathbf{I})] \in \left( \sigma_{\text{ML}}^2(n) - \beta_{\text{ML}}(n),\ \sigma_{\text{ML}}^2(n) + \beta_{\text{ML}}(n) \right),
\end{equation}
where the nominal variance $\sigma_{\text{ML}}^2(n)$ matches the CRLB:
\begin{equation}
\sigma_{\text{ML}}^2(n) = \sigma_{x_c}^2 = \left( \sum_{i=1}^n \frac{(F \cdot \frac{d g_i(x_c)}{d x_c})^2}{F g_i(x_c) + B_i} \right)^{-1}.
\end{equation}
As in the WLS case, the explicit expressions for $\epsilon_{\text{ML}}(n)$ and $\beta_{\text{ML}}(n)$ follow from the general bounds in \citet{espinosa2018optimality}, but they involve lengthy combinations of higher-order moments and derivatives of $g_i(x_c)$, and are therefore omitted for brevity.
\end{teo}
This result leads to a particularly important consequence: the nominal variance $\sigma^2_{\text{ML}}(n)$ coincides exactly with the CRLB (see Eq. (\ref{crbound})), making the ML estimator asymptotically optimal in this setting. Furthermore, the associated variance bound $\beta_{\text{ML}}(n)$ becomes negligible in the high-SNR regime (see Fig.~5 in \citet{espinosa2018optimality}), meaning that the variance of the ML estimator concentrates tightly around the CRLB. The bias also remains well-controlled via $\epsilon_{\text{ML}}(n)$, confirming that the ML estimator behaves as an approximately unbiased estimator even under realistic noise conditions. These results underscore the ML estimator's theoretical and practical advantages over alternative methods such as WLS, especially when dealing with Poisson-limited data and structured PSFs.

While the general theory in Theorem~\ref{espit} applies broadly, Theorem~\ref{espitd} demonstrates the special role of ML as the estimator that approaches the theoretical performance limit for astrometry.

\subsection{Adaptive weighted least squares estimator}
\label{awlssec}
As we mention before the WLS estimator depends critically on the choice of its weights. In light of Theorem \ref{espitdW}, the ideal weights should be proportional to the inverse of the expected intensity at each pixel, allowing the estimator to account for measurement uncertainties more effectively. However, since the true intensity distribution is unknown, a fixed weighting scheme often leads to biased or suboptimal results.

An adaptive weighted least squares (AWLS) approach has been proposed to address this issue. Instead of using pre-defined weights, the AWLS estimator dynamically adjusts weights based on the observed data, making it more responsive to varying observational conditions. This method improves estimation accuracy while maintaining the computational efficiency of WLS, making it an attractive alternative to ML estimator, which is optimal but computationally more expensive.

 \cite{espinosa2018optimality} discusses the development of an AWLS estimator designed to improve the performance of traditional WLS, particularly in high SNR regimes. However, since the true value of $x_c$ is unknown, this ideal weighting cannot be directly implemented. To overcome this limitation, the authors propose an adaptive WLS approach that dynamically adjusts the weights based on the observed data. Instead of using fixed weights, they define a data-driven weight selection scheme where the weight at each pixel is set as $\hat{w}_i(I_i) = \frac{1}{I_i}$, where $I_i$ is the observed intensity at that pixel. This approach makes $ \hat{w}_i$ a noisy but reasonable approximation of the optimal weight. Although this data-driven choice is a noisy approximation of the ideal $1/\lambda_i(x_c,F)$ weighting, it provides a practical alternative that avoids the need for prior knowledge of the true source parameters. In practice, this approximation replaces the unobservable expectation $\lambda_i(x_c,F)$ with the measured counts $I_i$, providing an implementable surrogate for the optimal weighting scheme.
 
 By incorporating this adaptive weighting scheme, the estimation problem is formulated as a minimization of the following modified cost function:

\begin{equation}
J_{\text{AWLS}}(\alpha, \mathbf{I}) = \sum_{i=1}^{n} \hat{w}_i(I_i) (I_i - \lambda_i(\alpha))^2.
\end{equation}

This formulation ensures that the estimator adapts to the observed pixel values, improving its performance in comparison to the traditional WLS estimator.

Importantly, the numerical analysis presented in \citet{espinosa2018optimality} shows that the adaptive WLS estimator significantly reduces the gap between WLS and ML estimators in terms of performance (see Fig.~7 therein). The MSE of the adaptive WLS estimator closely approaches the CRLB, making it a nearly optimal estimator while maintaining the computational efficiency of WLS. The results indicate that this method provides a practical and efficient alternative to ML estimator, particularly in scenarios where computational constraints are a concern.

The AWLS method is also closely related to weight selection schemes used in PSF fitting techniques, such as DAOPHOT (Dominion Astrophysical Observatory Photometry code, \citet{1987PASP...99..191S}). This similarity suggests that adaptive weighting techniques could be integrated into existing astronomical software to improve astrometric precision without significant computational overhead.

\subsection{Summary of estimators performance}

Table~\ref{tab:estimators_summary} summarizes key properties of the main estimators discussed in this work. It highlights their variance performance relative to the CRLB, robustness to low-SNR conditions, and whether they can achieve the CRLB.

\begin{table}[ht]
\centering
\caption{Summary of the relative performance and stability of different estimators under varying SNR conditions.}
\label{tab:estimators_summary}
\begin{tabular}{lccc}
\toprule
\textbf{Estimator} & \textbf{Variance vs. CRLB} & \multicolumn{2}{c}{\textbf{Performance Stability}} \\
\cmidrule(lr){3-4}
 &  & Low SNR & High SNR \\
\midrule
LS       & Suboptimal  & Yes        & No  \\
WLS        & Closer to CRLB   & Acceptable & No  \\
AWLS    & Optimal     & Yes        & Yes \\
ML & Optimal     & Yes        & Yes \\
\bottomrule
\end{tabular}
\end{table}
As shown in Table~\ref{tab:estimators_summary}, both the AWLS and ML estimators attain the CRLB across SNR regimes, confirming their statistical efficiency and adaptability. In contrast, LS and WLS remain limited by their fixed weighting schemes, which prevent them from adapting to signal-dependent noise variations, particularly in high-SNR conditions. The “Variance vs. CRLB” column summarizes the relative efficiency of each method, while the “Performance Stability” columns describe how consistently each estimator maintains precision when transitioning from low- to high-SNR regimes. Although WLS performs closer to the CRLB than LS due to its weighting, its fixed weights make it more sensitive to mismatches between the assumed and true noise models.  Conversely, the AWLS and ML estimators dynamically adjust to the data, ensuring near-optimal performance even under strong flux or background variations.

\section{Performance analysis and estimators in photometry} \label{sec:photo_estimators} 

In photometric analysis, estimating the flux of a point source while simultaneously determining the background level is a fundamental challenge. Traditional aperture photometry techniques often approach this task sequentially: first estimating the background flux from an annular region surrounding the source, and then subtracting this background contribution from the total flux measured within a circular aperture centered on the source. This sequential subtraction-based method, extensively described in \citet{howell1989two}, remains popular due to its simplicity and low computational cost. However, such techniques do not exploit the full statistical structure of the observational model.

Sequential estimation is suboptimal because it treats the background as a fixed quantity rather than as a parameter to be inferred jointly with the source flux. From an information-theoretic perspective, the data processing inequality \citep{cover1999elements} implies that such an approach leads to information loss, hindering the ability of any estimator to attain the CRLB. 

To address this, \citet{vicuña} propose a joint estimation framework in which both the source flux and background are estimated simultaneously from all pixels in the region of interest. This joint modeling approach leads to improved accuracy, tighter variance bounds, and lower bias in the resulting flux estimates. Their study introduces new theoretical bounds on estimator performance, validated through both numerical simulations and real-world photometric data from the Transiting Exoplanet Survey Satellite (TESS). In what follows, we present the main theoretical result that characterizes these bounds in the context of joint estimation of flux and background.

\subsection{Bounding the performance of joint estimators}

One of the key theoretical contributions of the work by \citet{vicuña} is Theorem~\ref{mariot}, which generalizes the results of \citet{espinosa2018optimality} to multivariate settings relevant to photometry. This result establishes bias and variance bounds for estimators that jointly infer multiple parameters—most notably, the source flux $F$ and the background level $B$—within a unified optimization framework.

\begin{teo}\citep[Theorem 4.1]{vicuña} 
\label{mariot} 
Let $\boldsymbol{\alpha}$ be the true parameter vector representing the source and background fluxes, and let $\tau_J(\mathbf{I})$ be the estimator defined as the solution to an optimization problem. Then, for each component $j \in \{1, ..., p\}$, there exists constants $\epsilon_{J,j}(n)$, $\sigma^2_{J,j}(n)$, and $\beta_{J,j}(n)$ such that:
\begin{equation}
    \left| \mathbb{E} \{ \tau_{J,j}(\mathbf{I}) \} - \alpha_j \right| \leq \epsilon_{J,j}(n),
\end{equation}
\begin{equation}
    Var \{ \tau_{J,j}(\mathbf{I}) \} \in \left( \sigma^2_{J,j}(n) - \beta_{J,j}(n),\ \sigma^2_{J,j}(n) + \beta_{J,j}(n) \right).
\end{equation}
Here, $\sigma^2_{J,j}(n)$ denotes the nominal variance of the estimator associated to the parameter $\alpha_j(n)$, while $\beta_{J,j}(n)$ bounds the deviation from this nominal value\footnote{Explicit formulas for $\sigma^2_{J,j}(n)$, $\beta_{J,j}(n)$, and $\epsilon_J(n)$ can be found in \citet{vicuña}.}.
\end{teo}

Theorem~\ref{mariot} shows suitable regularity conditions, where joint estimators are nearly unbiased and exhibit variances close to a nominal quantity. As in the work of \citet{espinosa2018optimality}, the tightness of the variance bounds $\beta_{J,j}(n)$ is strongly influenced by the SNR. In the high-SNR regime, $\beta_{J,j}(n)$ becomes negligible, allowing the estimator's variance to concentrate sharply around $\sigma^2_{J,j}(n)$—an analog to the asymptotic efficiency observed for the ML estimator in astrometry. This reinforces the interpretation of high SNR as a condition under which optimal performance is practically attainable.

Importantly, the theorem's generality makes it applicable to several estimators, including WLS, AWLS, and ML. It thus provides a unified analytical framework for comparing their performance in realistic observational settings. Notably, in the case of the ML estimator, the nominal variance $\sigma^2_{J,j}$ exactly matches the CRLB under standard regularity conditions. Consequently, the ML estimator retains its theoretical optimality in this photometric context: not only does it achieve the CRLB, but its practical performance aligns remarkably well with these bounds. As shown by \citet{vicuña}, the ML estimator exhibits negligible bias and near-minimal variance across a wide range of observational conditions.

\subsection{Validation with real observational data}

To assess the practical relevance of some relevant estimators in the context of photometry, \citet{vicuña} conducted a comprehensive numerical and observational validation. Simulated photometric datasets were used to compare WLS, AWLS, and ML estimators under varying noise conditions and source brightness levels. The results confirm that AWLS significantly outperforms traditional WLS by adapting pixel weights to local noise statistics, effectively reducing both bias and variance. As expected, the ML estimator consistently achieves the lowest possible variance, validating its theoretical optimality in photometry.

These findings were further corroborated using real photometric data from TESS \citep{ricker2015tess}. The study demonstrated that joint estimation via ML outperforms traditional sequential estimation methods, particularly in crowded or low-SNR settings. Incorporating background estimation directly within the optimization process led to significantly more accurate flux recovery and reduced systematic errors.

To summarize, the theoretical and empirical results presented in \citet{vicuña} provide strong evidence that joint estimation strategies outperform traditional sequential techniques in photometric analysis. On the technical side, Theorem~\ref{mariot} offers a principled foundation for evaluating estimator performance, while the experimental results underscore the practical advantages of ML and AWLS. As a result, modern photometric pipelines should consider incorporating joint estimation frameworks to approach the performance limits dictated by the CRLB.

\section{Extensions and Applications in Astrometry and Photometry} \label{sec:extensions}

Astrometric modeling continues to evolve in response to increasingly complex observational challenges. While the classical CRLB provides a powerful tool for understanding the limits of estimation under a frequentist framework, it relies on the assumption that the parameters to be estimated are deterministic but unknown. In practical scenarios, however, additional information—such as prior measurements or dynamical models—can be available and should be incorporated into the estimation process. 

This section explores two important extensions of the classical model presented in Section \ref{sec:model}. First, we examine the Bayesian Cramér–Rao lower bound (BCRLB), which accounts for prior uncertainty in parameter estimation. This strategy is particularly effective in low-SNR regimes or when prior knowledge in the form of catalogs are available. Second, we briefly discuss astrometric modeling for moving sources, which introduces additional complexity due to temporal evolution and trajectory fitting.

\subsection{A Bayesian approach}

\citet{echeverria2016analysis} investigates the role of prior information in the estimation of astrometric positions by analyzing the BCRLB. Unlike the classical CRLB, which provides a lower bound on the variance of any unbiased estimator without incorporating prior knowledge, the BCRLB is formulated within the Bayesian framework \citep{vantrees1968}. It quantifies a lower bound on the MSE of any estimator by taking into account both the likelihood information from the data and the prior distribution of the parameters \citep{vantrees2007}.

Importantly, the BCRLB is not a generalization of the CRLB but rather its Bayesian counterpart. It provides the fundamental performance limit for estimators when the parameters are treated as random variables with a known prior. The BCRLB is thus particularly relevant in contexts where reliable prior information is available in the estimation task, such as from earlier catalogs or models based on orbital dynamics.

The results in \citet{echeverria2016analysis} demonstrate that the BCRLB can be substantially tighter than the CRLB, especially under low SNR conditions or when the detector resolution is coarse. Moreover, the authors show that the conditional expectation \citep{kay1993fundamentals} —which minimizes the Bayesian risk under squared error loss—nearly attains the BCRLB. This makes the conditional expectation a highly attractive estimator in practical Bayesian astrometry, offering both theoretical optimality and robustness to observational limitations.

This desirable behavior is formally supported by the following result, which establishes a lower bound on the MSE achievable by any estimator in the Bayesian setting.
\begin{teo}\citep[Theorem 2]{echeverria2016analysis}
\label{alext}
For any estimator $\tau_n : \mathbb{N}^n \rightarrow \mathbb{R}$, its MSE satisfies the following inequality:
\begin{equation}
\mathbb{E}_{X_c,  \mathbf{I}} \left\{ \left( \tau_n( \mathbf{I}) - X_c \right)^2 \right\} 
\geq \left[ \mathbb{E}_{X_c,  \mathbf{I}} \left\{ 
\left( \frac{d}{dx} \ln \tilde{L}(\mathbf{I},X_c) \right)^2 
\right\} \right]^{-1},
\label{eq:bcr}
\end{equation}
where $ \tilde{L}(\mathbf{I},X_c) = L(\mathbf{I}; x_c, F) \cdot \psi(x_c) $ is the joint density of the data $\mathbf{I}$ and $ \psi(x_c) $ is the prior density of the random variable $X_c $.

The bound in Eq. (\ref{eq:bcr}) leads to the following expression for the Bayesian Fisher Information (BFI):
\begin{equation}
\text{BFI}(F, \psi) = \mathbb{E}_{X_c \sim \psi} \left \{ \mathcal{I}_{X_c} \right\} 
+ \mathbb{E}_{X_c \sim \psi} \left\{ 
\left( \frac{d}{dx} \ln \psi(X_c) \right)^2 
\right\},
\label{addi}
\end{equation}
where the first term is the expected classical Fisher information from the observations, and the second term represents the information contained in the prior.
\end{teo}

Theorem \ref{alext} highlights how prior knowledge can be mathematically integrated into the estimation process to yield a more informative and tighter lower bound on estimation error. Unlike the classical CRLB, which solely depends on the likelihood function derived from the data, the Bayesian version accounts for additional knowledge through the prior distribution $\psi$, which is the additive term presented in Eq. (\ref{addi}). 

This result has particular importance in the field of astrometry, where catalogs and physical models often provide reliable prior information about the probable location of sources.

Numerical results presented by \citet{echeverria2016analysis} confirm that in scenarios with low SNR or coarse pixel resolution, where the classical CRLB provides weak guarantees, the Bayesian estimator, informed by prior knowledge, maintains high accuracy and precision. Importantly, in these cases the conditional expectation offers an implementable solution that achieves the performance gains represented by the BCRLB in Eq. (\ref{eq:bcr}).

\subsection{Moving sources}

Accurate astrometric measurements of celestial objects depend critically on precise centroid estimation, particularly when the source exhibits motion during the exposure. \citet{bouquillon2017characterizing} analyze the fundamental precision limits in the astrometric localization of moving point sources, extending prior results for one-dimensional detectors to the more realistic case of two-dimensional digital sensor arrays. 

In their formulation, the PSF of a moving source is modeled as the convolution of a stationary PSF with the trajectory induced by linear motion (see Fig. \ref{fig:moving}). This formulation of a dynamic PSF leads to closed-form expressions for the CRLB, which quantify the minimum achievable variance in positional estimates under varying observational parameters .

\begin{figure}[ht]
    \centering
    \includegraphics[width=0.6\textwidth]{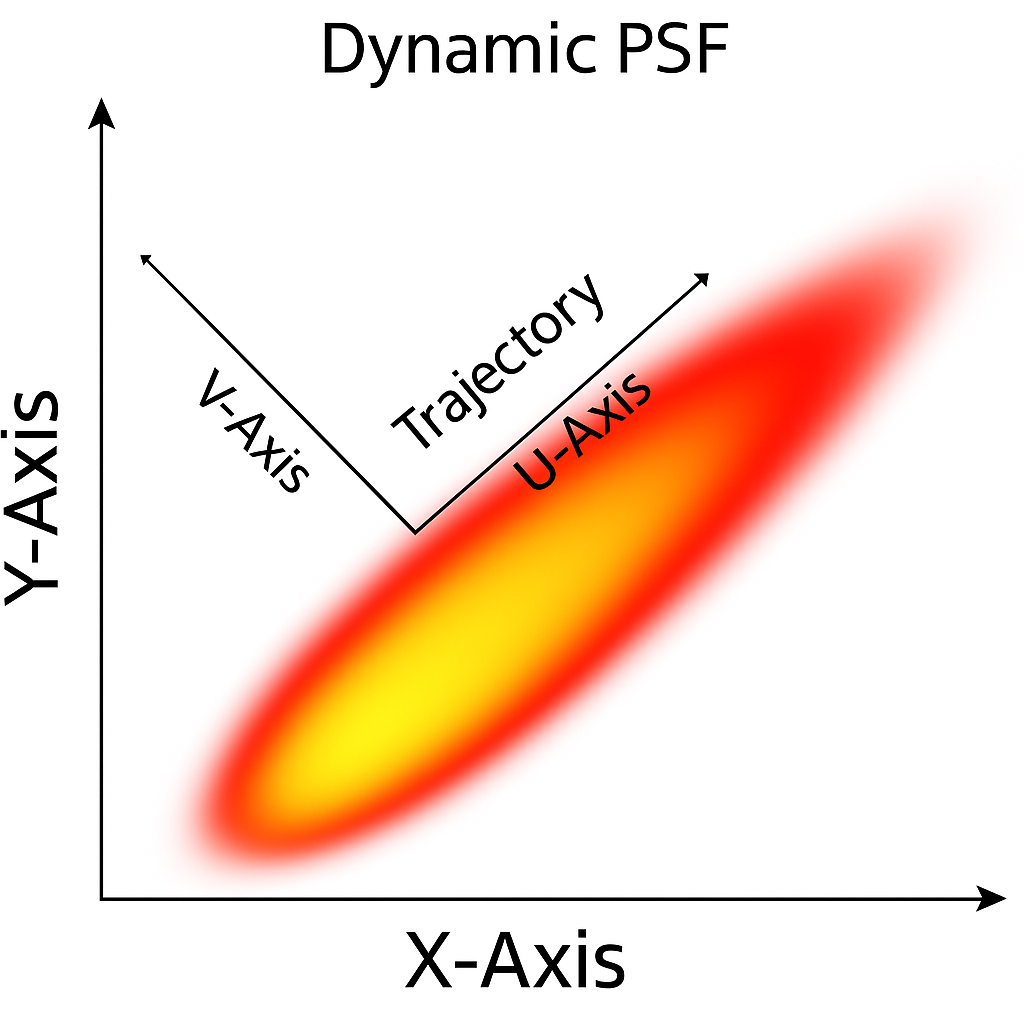}
    \caption{Illustration of a dynamic PSF for a moving source. The PSF is elongated along the trajectory direction due to the linear motion of the source during the exposure. The color gradient indicates intensity, with warmer colors representing higher flux levels. This schematic is inspired by \citet{bouquillon2017characterizing} but adapted for illustrative purposes.}
    \label{fig:moving}
\end{figure}

For a source moving along the detector plane and imaged with a 
two-dimensional array, the Cramér–Rao Lower Bound (CRLB) for the
astrometric precision along the drifting direction $U$ can be written as
\begin{equation}
\sigma^2_{\mathrm{CRU}} =
\frac{4\pi^2\sigma^4 L^2}{F^2}
\left[
\sum_{j=1}^{n_y}\sum_{i=1}^{n_x}
\frac{(N_i^j)^2}{D_i^j}
\right]^{-1},
\label{eq:CRLB-moving-2D}
\end{equation}
where $F$ is the total source flux, $\sigma$ is the width of the
instantaneous Gaussian PSF, and $L$ is the drift length during the exposure. The terms $N_i^j$ and $D_i^j$ represent, respectively, the sensitivity of the pixel-integrated signal to small displacements of the centroid along $U$, and the expected total counts in each pixel, including the contributions from the sky background, detector components (dark current and readout noise, treated in their effective Poisson form), and the pixel-integrated flux of the moving source.

Closed-form expressions for $N_i^j$ and $D_i^j$—including their dependence on the rotated $(U,V)$ coordinates and on the motion kernel—are provided in \citet[][Sect.~4.3.1]{bouquillon2017characterizing},
to which we refer the reader for the full derivation.

This CRLB-driven formulation reveals key trade-offs in the precision of astrometric measurements for moving objects. Unlike stationary sources—where longer exposures invariably improve precision—in the moving case, extended exposure times lead to flux spreading across pixels. This spreading reduces the SNR per pixel and increases sensitivity to background contamination, ultimately degrading positional accuracy.

One of the principal practical implications of this analysis is the existence of an optimal exposure time, which minimizes the CRLB by balancing increased photon collection against the detrimental effects of motion blur. In the over- and well-sampled regimes, the CRLB curve exhibits a characteristic U-shape: it initially decreases with exposure duration, reaches a minimum, and then increases again as motion dominates. For faint, sky-limited sources this minimum typically occurs when the drift length is comparable to the image quality ($L \approx \mathrm{FWHM}$). Its location also depends more generally on the object’s speed, the PSF width, detector resolution, and the level of background and readout noise.

\citet{bouquillon2017characterizing} further quantify the impact of motion on achievable precision. In ideal low-noise conditions and using the optimal exposure time, the CRLB degrades by about 19–33\% relative to the stationary case; when readout or background noise is significant, this degradation can increase slightly. In the undersampled regime, additional oscillations appear due to discretization effects, but these do not alter the general conclusion that an optimal exposure exists.

The analytical expressions derived can therefore guide the design of astrometric instruments and inform the development of tracking algorithms that dynamically adjust exposure based on expected motion profiles, while also providing simple criteria (such as the $L \approx \mathrm{FWHM}$ rule of thumb) for planning efficient observations of moving targets.

\subsection{Applications in Modern Surveys and Photometric Pipelines}

The theoretical framework developed throughout this review—based on the FIM, the CRLB, and optimal strategies for flux and position estimation—has become increasingly relevant for the analysis and design of modern astronomical surveys. Over the past two decades, the interplay between theoretical precision limits and operational data-reduction pipelines has evolved into a cornerstone of both astrometric and photometric science. Contemporary missions, whether space- or ground-based, now routinely achieve precision levels that are only a small factor above their information-theoretic limits, underscoring the practical importance of the methods discussed here.

An emblematic example is the Gaia mission, whose astrometric solution in Data Release~2 demonstrates that, for well-calibrated sources, positional uncertainties approach the sub-milliarcsecond regime. The centroiding and flux-extraction processes in Gaia rely on PSF and LSF fitting procedures that implicitly embody Fisher-information principles. The formal errors reported in the catalogues correspond closely to CRLB predictions given the photon statistics, read noise, and pixel sampling of the instrument \citep{lindegren2018gaia}. A similar connection emerges in the treatment of systematic effects such as charge-transfer inefficiency and attitude jitter, which effectively shift practical performance away from the theoretical bounds derived under idealized conditions.

Ground-based wide-field surveys such as LSST and Pan-STARRS face complementary challenges, where atmospheric transparency, spatially variable PSFs, and detector non-uniformities dominate the information budget. The pioneering work of \citet{stubbs2007toward} demonstrated that achieving millimagnitude-level photometric uniformity requires direct characterization of the optical transmission function of both atmosphere and instrument, linking physical calibration to the Fisher-information structure underlying the data. Likewise, the comparative analysis by \citet{becker2007pursuit} on photometric pipelines for LSST illustrates how algorithmic choices—aperture integration, PSF fitting, or deblending strategies—correspond to different operating points within the precision trade-offs established by the CRLB.

The analysis of astrometric precision tests on TESS data \citep{gai2022astrometric} confirmed that the empirical dependence of centroiding precision on the inverse of the signal-to-noise ratio closely follows the theoretical behavior predicted by information-limited models. Although the Fisher formalism is not explicitly invoked, their results demonstrate that, under realistic noise and sampling conditions, TESS measurements approach the photon-limited regime consistent with Cramér–Rao expectations.

The same information-theoretic principles have also inspired recent methodological developments beyond specific missions. For instance, \citet{espinosa2025impact} revisited the classical problem of aperture optimization in photometry—originally treated in terms of SNR by \citet{howell1989two}—from a Fisher-information perspective. Their analysis showed that SNR-based criteria can lead to suboptimal precision when flux and background must be estimated jointly, while a Fisher-based formulation yields apertures whose variance asymptotically approaches the Cramér–Rao limit. This result provides a unified statistical foundation for aperture selection and exemplifies how the Fisher-information framework continues to guide both the interpretation and design of high precision photometric pipelines.

\section{Conclusions}
\label{sec:conclusions}
In this review, we have analyzed the theoretical foundations and practical implementations of key estimators for astrometry and photometry of celestial point sources. Our discussion traced the transition from empirical centroiding methods to rigorous pixel-level statistical models, highlighting how estimation theory—and especially the CRLB—clarifies the fundamental precision limits attainable.

We showed that estimators such as the ML can asymptotically attain the CRLB under specific conditions, while others like the LS estimator remain suboptimal in high-SNR regimes. Recent developments have demonstrated that weighted and adaptive least-squares methods (e.g., AWLS) significantly reduce both bias and variance in practical applications.

Importantly, two key extensions of the CRLB framework have broadened its applicability. First, the Bayesian variant of the CRLB (BCRLB) allows the incorporation of prior knowledge, leading to tighter performance bounds—particularly valuable in low SNR scenarios or when catalog information is available. Second, models for tracking moving sources have generalized the CRLB to account for motion-induced PSF distortions, revealing fundamental trade-offs between exposure time and localization accuracy.

These findings collectively support the systematic incorporation of optimal inference tools in observational pipelines and the use of advanced statistical bounds to guide estimator design, especially in the context of next-generation astronomical surveys.

\subsection{Future work}

While the theoretical bounds and estimator performance reviewed in this survey provide a solid foundation and a up-to-date perspective for astrometric and photometric precision, several challenges persist—particularly in the face of rapidly evolving observational conditions and data acquisition technologies. New astronomical environments demand the development of more robust, adaptive, and scalable estimation strategies.

One major concern is the growing impact of satellite mega-constellations in low Earth orbit, which has introduced a new source of transient noise in the form of streak contamination across short-exposure images. This contamination can significantly degrade centroid precision and flux estimation, especially in wide-field surveys during twilight hours~\citep{lawrence2022}. Addressing this requires not only effective image-cleaning procedures, but also the development of statistical models that explicitly account for non-stationary, time-dependent background interference. Incorporating spatio-temporal priors or robust estimators capable of identifying and suppressing satellite streaks will be essential in future data reduction pipelines.

At the same time, upcoming facilities such as the Vera C. Rubin Observatory (LSST) and Euclid will produce data at unprecedented volume and cadence, making computational efficiency a central requirement. In this context, there is growing interest in fast, information-aware estimators that remain close to optimality. Stochastic or adaptive versions of WLS, for instance, offer a promising compromise between performance and tractability~\citep{espinosa2018optimality}. Future efforts should focus on further improving these estimators by integrating dimensionality reduction techniques, approximate inference schemes, and online learning to meet the demands of real-time survey operations.

The integration of prior information into the estimation process also remains a powerful yet underutilized direction. Bayesian approaches informed by stellar catalogs, proper motion models, or prior instrument calibrations can substantially improve performance in low-SNR regimes. However, the full potential of the BCRLB in guiding the design of such estimators—particularly under structured priors and model uncertainty—warrants further investigation~\citep{echeverria2016analysis}. Extending current theoretical tools to support inference in high-dimensional and hierarchical Bayesian models could lead to new advances in position and flux recovery.

Finally, future work should explore how CRLB-based frameworks can be generalized to address increasingly complex observational conditions. This includes variable and spatially structured PSFs, irregular detector geometries, and motion-induced distortions. For example, tracking fast-moving sources like near-Earth objects introduces an additional layer of difficulty: the trade-off between motion blur and exposure time must be quantitatively modeled to determine the optimal observation strategy. Recent results~\citep{bouquillon2017characterizing} have extended the CRLB to account for such dynamics, but further work is needed to integrate these formulations into practical astrometric algorithms.

The next phase of development in astrometric and photometric inference will require not only refining estimators but also adapting them to the new observational frontier—where satellite contamination, big data constraints, and complex source behavior converge. Advancing along these lines will demand deeper synergy between statistical methodology and astronomical instrumentation, with estimation theory continuing to serve as a guiding framework.

\section{Glossary}

\begin{description}

  \item[Adaptive Weighted Least Squares Estimation (AWLS):] A variation of WLS that incorporates stochastic modeling of the weights or the noise, often yielding more accurate estimates under certain observational regimes.

    \item[Cramér-Rao Lower Bound (CRLB):] A theoretical lower bound on the variance of an unbiased estimator, providing a benchmark for estimator performance.

\item[Bayesian Cramér-Rao Lower Bound (BCRLB):] The Bayesian counterpart of the CRLB, which provides a lower bound on the mean squared error of any estimator when prior information about the parameters is incorporated through a prior distribution.

  \item[Charge-Coupled Device (CCD):] A type of electronic detector widely used in astronomy to record images. CCDs convert incoming photons into electrons, which are then read out and digitized, providing high sensitivity, linear response, and low noise compared to photographic plates.

 \item[Full-Width at Half-Maximum (FWHM):] A measure of the width of a distribution or function, defined as the distance between the two points where the function reaches half of its maximum value. The FWHM of the PSF is commonly used as a proxy for image resolution and seeing conditions.
 
  \item[Least Squares Estimation (LS):] A method for parameter estimation by minimizing the sum of the squared differences between observed and modeled data.
  
  \item[Maximum Likelihood Estimation (ML):] A statistical method that estimates parameters by maximizing the likelihood function given the observed data.

  \item[Point Spread Function (PSF):] The response of an imaging system to a point source, representing how the light from a single point is distributed across the detector.
  
  \item[Mean Squared Error (MSE):] The MSE of an estimator $\hat{\theta}$ for a parameter $\theta$ is defined as $MSE(\hat{\theta}) = \mathbb{E}\left[(\hat{\theta} - \theta)^2\right].$ When $\hat{\theta}$ is unbiased, then $MSE(\hat{\theta})=Var[\hat{\theta}]$

  \item[Signal-to-Noise Ratio (SNR):] A dimensionless measure expressing the strength of the desired signal relative to the background noise. High SNR typically implies better measurement precision.

  \item[Weighted Least Squares Estimation (WLS):] A generalization of least squares where each data point contributes proportionally to a specified weight, often used to account for heteroskedastic noise.

\end{description}
\section{Funding sources}
SAET, JFS, and MO acknowledge support from the Advanced Center for Electrical and Electronic Engineering (AC3E), funded by the Chilean ANID Basal Project (CIA 250006). Additionally, JFS acknowledges support from Fondecyt Project 1250098, MO acknowledges support from Fondecyt Project 1250036. RAM acknowledges support from FONDECYT/ANID grant \# 1240049 and from ANID, Fondo GEMINI, Astrónomo de Soporte GEMINI-ANID grant \# 3223 AS0002.

\section{Declaration of generative AI and AI-assisted technologies in the writing process.}
During the preparation of this work the author(s) used ChatGPT (OpenAI) and Grammarly in order to support the writing process, specifically to improve the readability and language of the manuscript. After using this tool, the author(s) reviewed and edited the content as needed and take(s) full responsibility for the content of the published article.

\bibliographystyle{elsarticle-harv} 
\bibliography{referencesb}

@article{mendez2013,
         author={Mendez, René A. and Silva, Jorge F. and Lobos, Rodrigo},
         title={Analysis and Interpretation of the {C}ramér-{R}ao Lower-Bound in Astrometry: One-Dimensional Case},
         journal={Publications of the Astronomical Society of the Pacific},
         year={2013},
         volume={125},
         number={927},
         pages={580-594},
         month={May},
}

@article{mendez2014,
         author={Mendez, René A. and Silva, Jorge F. and Oróstica, Rodrigo and Lobos, Rodrigo},
         title={Analysis of the {C}ramér-{R}ao Bound in the Joint Estimation of Astrometry and Photometry},
         journal={Publications of the Astronomical Society of the Pacific},
         year={2014},
         volume={126},
         number={942},
         pages={798-810},
         month={August},
}

@ARTICLE{vicuña,
       author = {{Vicu{\~n}a}, Mario L. and {Silva}, Jorge F. and {Mendez}, Rene A. and {Orchard}, Marcos E. and {Espinosa}, Sebastian and {Tregloan-Reed}, Jeremy},
        title = "{Optimal Photometry of Point Sources: Joint Source Flux and Background Determination on Array Detectors-from Theory to Practical Implementation}",
      journal = {Publications of the Astronomical Society of the Pacific},
         year = 2024,
        month = jan,
       volume = {136},
       number = {1},
          eid = {014501},
        pages = {014501},
          doi = {10.1088/1538-3873/ad0ca3},
archivePrefix = {arXiv},
       eprint = {2311.12142},
 primaryClass = {astro-ph.IM},
       adsurl = {https://ui.adsabs.harvard.edu/abs/2024PASP..136a4501V}
}

@article{anderson2000toward,
  title={Toward high-precision astrometry with WFPC2. I. Deriving an accurate point-spread function},
  author={Anderson, Jay and King, Ivan R},
  journal={Publications of the Astronomical Society of the Pacific},
  volume={112},
  number={776},
  pages={1360},
  year={2000},
  publisher={IOP Publishing}
}

@book{cover1999elements,
  title={Elements of information theory},
  author={Cover, Thomas M},
  year={1999},
  publisher={John Wiley \& Sons}
}

@article{bouquillon2017characterizing,
  title={Characterizing the astrometric precision limit for moving targets observed with digital-array detectors},
  author={Bouquillon, S{\'e}bastien and Mendez, RA and Altmann, Martin and Carlucci, Teddy and Barache, Christophe and Taris, F and Andrei, AH and Smart, R},
  journal={Astronomy \& Astrophysics},
  volume={606},
  pages={A27},
  year={2017},
  publisher={EDP Sciences}
}

@article{echeverria2016analysis,
  title={Analysis of the Bayesian Cram{\'e}r-Rao lower bound in astrometry-Studying the impact of prior information in the location of an object},
  author={Echeverria, Alex and Silva, Jorge F and Mendez, Rene A and Orchard, Marcos},
  journal={Astronomy \& Astrophysics},
  volume={594},
  pages={A111},
  year={2016},
  publisher={EDP Sciences}
}

@article{espinosa2018optimality,
  title={Optimality of the maximum likelihood estimator in astrometry},
  author={Espinosa, Sebastian and Silva, Jorge F and Mendez, Rene A and Lobos, Rodrigo and Orchard, Marcos},
  journal={Astronomy \& Astrophysics},
  volume={616},
  pages={A95},
  year={2018},
  publisher={EDP Sciences}
}

@article{howell1989two,
  title={Two-dimensional aperture photometry: signal-to-noise ratio of point-source observations and optimal data-extraction techniques},
  author={Howell, Steve B},
  journal={Publications of the Astronomical Society of the Pacific},
  volume={101},
  number={640},
  pages={616},
  year={1989},
  publisher={IOP Publishing}
}

@article{rao1945,
         author={Rao, Calyampudi Radakrishna},
         title={Information and the Accuracy Attainable in the Estimation of Statistical Parameters},
         journal={Bulletin of the Calcutta Mathematical Society},
         year={1945},
         volume={37},
         number={3},
         pages={81-89}
}

@article{cramer1946,
author = { Harald   Cramér },
title = {A contribution to the theory of statistical estimation},
journal = {Scandinavian Actuarial Journal},
volume = {1946},
number = {1},
year  = {1946},
publisher = {Taylor & Francis},
doi = {10.1080/03461238.1946.10419631},

URL = { 
        https://doi.org/10.1080/03461238.1946.10419631
    
},
pages = {85-94},

}

@techreport{lindegren2008,
            author={Lindegren, L.},
            title={A general {M}aximum-{L}ikelihood algorithm for model fitting to CCD sample data},
            institution={Lund Observatory}, number={GAIA-C3-TN-LU-LL-078-01},
            year={2008},
            month={November}
}

@article{lobos2015,
         author={Lobos, Rodrigo A. and Silva, Jorge F. and Mendez, René A. and Orchard, Marcos},
         title={Performance Analysis of the Least-Squares Estimator in Astrometry},
         journal={Publications of the Astronomical Society of the Pacific},
         year={2015},
         volume={127},
         number={957},
        pages={1166 - 1182},
        month={November}
}

@book{vanaltena1986,
  author    = {Van Altena, William F. and Lee, Jae-Woo and Hoffleit, Dorrit},
  title     = {The General Catalogue of Trigonometric Stellar Parallaxes},
  publisher = {Yale University Observatory},
  year      = {1986}
}

@inproceedings{lindegren1978,
  title={Photoelectric astrometry-A comparison of methods for precise image location},
  author={Lindegren, Lennart},
  booktitle={IAU Colloq. 48: Modern Astrometry},
  pages={197},
  year={1978}
}

@inproceedings{adorf1995,
  author    = {Adorf, Hans-Martin},
  title     = {Photometry with the Hubble Space Telescope: Likelihood Methods and Limitations},
  booktitle = {Astronomical Data Analysis Software and Systems IV},
  series    = {ASP Conference Series},
  volume    = {77},
  pages     = {305--308},
  year      = {1995},
  editor    = {Shaw, R. A. and Payne, H. E. and Hayes, J. J. E.}
}

@inproceedings{adorf1996limits,
  title={Limits to the precision of joint flux and position measurements on array data},
  author={Adorf, Hans-Martin},
  booktitle={Astronomical Data Analysis Software and Systems V},
  volume={101},
  pages={13},
  year={1996}
}

@article{jakobsen1992cramer,
  title={The cramer-rao lower bound and stellar photometry with aberrated hst images},
  author={Jakobsen, P and Greenfield, P and Jedrzejewski, R},
  journal={Astronomy and Astrophysics (ISSN 0004-6361), vol. 253, no. 1, Jan. 1992, p. 329-332.},
  volume={253},
  pages={329--332},
  year={1992}
}

@article{1948JRSS...Anscombe,
    author = {Anscombe, F. J.},
    title = {The transformation of Poisson, binomial and negative-binomial data},
    journal = {Journal of the Royal Statistical Society. Series B (Methodological)},
    year = {1948},
    volume = {10},
    number = {2},
    pages = {192--242}
}

@article{1995Murtagh,
    author = {Murtagh, F. and Starck, J.-L. and Bijaoui, A.},
    title = {Image restoration with noise suppression using a multiresolution support},
    journal = {Astronomy and Astrophysics Supplement Series},
    year = {1995},
    volume = {112},
    pages = {179--189}
}

@article{lawrence2022,
  title={Impact of satellite constellations on optical astronomy and recommendations toward mitigation},
  author={Lawrence, Andy and others},
  journal={Nature Astronomy},
  volume={6},
  pages={130–136},
  year={2022},
  doi={10.1038/s41550-022-01777-6}
}

@article{perryman1997hipparcos,
  title={The HIPPARCOS catalogue},
  author={Perryman, Michael AC and Lindegren, L and Kovalevsky, J and Hoeg, E and Bastian, U and Bernacca, PL and Cr{\'e}z{\'e}, M and Donati, F and Grenon, M and Grewing, M and others},
  journal={Astronomy and Astrophysics, Vol. 323, p. L49-L52},
  volume={323},
  pages={L49--L52},
  year={1997}
}

@article{gaia2016,
  title={The gaia mission},
  author={Prusti, Timo and De Bruijne, JHJ and Brown, Anthony GA and Vallenari, Antonella and Babusiaux, C and Bailer-Jones, CAL and Bastian, U and Biermann, M and Evans, Dafydd Wyn and Eyer, L and others},
  journal={Astronomy \& astrophysics},
  volume={595},
  pages={A1},
  year={2016},
  publisher={EDP sciences}
}

@book{hoskin1997hipparchus,
  title={The history of Astronomy: a very short Introduction},
  author={Hoskin, Michael},
  year={2003},
  publisher={OUP Oxford}
}

@article{mignard2018gaia,
  title={Gaia data release 2-the celestial reference frame (gaia-crf2)},
  author={Mignard, Fran{\c{c}}ois and Klioner, Sergei A and Lindegren, L and Hernandez, J and Bastian, U and Bombrun, A and Hobbs, Dick and Lammers, Uwe and Michalik, Dirk and Ramos-Lerate, Mercedes and others},
  journal={Astronomy \& astrophysics},
  volume={616},
  pages={A14},
  year={2018},
  publisher={EDP sciences}
}

@article{sozzetti2005,
  title={Astrometric methods and instrumentation to detect and characterize extrasolar planets: a review},
  author={Sozzetti, A.},
  journal={Publications of the Astronomical Society of the Pacific},
  volume={117},
  pages={1021--1048},
  year={2005}
}

@book{kay1993fundamentals,
  title={Fundamentals of Statistical Signal Processing: Estimation Theory},
  author={Kay, Steven M.},
  year={1993},
  edition={1st},
  publisher={Prentice Hall}
}

@article{gai2017performance,
  title={Performance of an algorithm for estimation of flux, background, and location on one-dimensional signals},
  author={Gai, Mario and Busonero, Deborah and Cancelliere, Rossella},
  journal={Publications of the Astronomical Society of the Pacific},
  volume={129},
  number={975},
  pages={054502},
  year={2017},
  publisher={IOP Publishing}
}

@book{lupton1993statistics,
  title={Statistics in Theory and Practice},
  author={Lupton, R.},
  year={1993},
  publisher={Princeton University Press}
}

@article{2001sccd.book.....J,
  title={Scientific charge-coupled devices},
  author={Janesick, James R and Elliott, Tom and Collins, Stewart and Blouke, Morley M and Freeman, Jack},
  journal={Optical Engineering},
  volume={26},
  number={8},
  pages={692--714},
  year={1987},
  publisher={Spie}
}

@BOOK{2006hca..book.....H,
       author = {{Howell}, Steve Bruce},
        title = "{Handbook of CCD Astronomy}",
         year = 2006,
       volume = {5},
       adsurl = {https://ui.adsabs.harvard.edu/abs/2006hca..book.....H}
}

@book{2007ptd..book.....J,
  title={Photon transfer},
  author={Janesick, James R},
  number={PUBDB-2021-04195},
  year={2007},
  publisher={SPIE press}
}

@BOOK{2008eiad.book.....M,
  author = {{McLean}, Ian S.},
  title = "{Electronic Imaging in Astronomy: Detectors and Instrumentation (Second Edition)}",
  year = 2008,
  publisher = {Springer},
  adsurl = {https://ui.adsabs.harvard.edu/abs/2008eiad.book.....M}
}

@book{walter2000astrometry,
  title={Astrometry of fundamental catalogues: the evolution from optical to radio reference frames},
  author={Walter, Hans G and Sovers, Ojars},
  year={2000},
  publisher={Springer Science \& Business Media}
}

@book{seidelmann1992explanatory,
  title={Explanatory supplement to the astronomical almanac},
  author={Seidelmann, P Kenneth},
  year={1992},
  publisher={University Science Books}
}

@inproceedings{vanaltena1975,
  title={Digital image centering, {I}},
  author={Van Altena, William F. and Auer, L.H.},
  booktitle={Image Processing Techniques in Astronomy: Proceedings of a Conference Held in Utrecht on March 25--27, 1975},
  pages={411--418},
  year={1975},
  organization={Springer}
}

@article{winick1986,
  author    = {Kim A. Winick},
  title     = {Cramér-Rao Lower Bounds on the Performance of Charge-Coupled Device Optical Position Estimators},
  journal   = {Journal of the Optical Society of America A},
  volume    = {3},
  number    = {11},
  pages     = {1809--1815},
  year      = {1986},
  doi       = {10.1364/JOSAA.3.001809}
}

@techreport{chen1987,
  author      = {Yeunung Chen},
  title       = {Cramér-Rao Bounds on the Accuracy of Location and Velocity Estimations Using CCD Optical Sensors},
  institution = {MIT Lincoln Laboratory},
  year        = {1987},
  number      = {Technical Report 777},
  note        = {Available via the Defense Technical Information Center (DTIC), Accession Number ADA188481},
  url         = {https://apps.dtic.mil/sti/citations/ADA188481}
}

@book{vantrees1968,
  author    = {Harry L. Van Trees},
  title     = {Detection, Estimation, and Modulation Theory, Part I},
  year      = {1968},
  publisher = {John Wiley \& Sons},
  address   = {New York}
}

@book{vantrees2007,
  author    = {Harry L. Van Trees and Kristine L. Bell},
  title     = {Bayesian Bounds for Parameter Estimation and Nonlinear Filtering/Tracking},
  year      = {2007},
  publisher = {John Wiley \& Sons / IEEE Press},
  address   = {Piscataway, NJ}
}

@ARTICLE{1987PASP...99..191S,
       author = {{Stetson}, Peter B.},
        title = "{DAOPHOT: A Computer Program for Crowded-Field Stellar Photometry}",
      journal = {Publications of the Astronomical Society of the Pacific},
         year = 1987,
        month = mar,
       volume = {99},
        pages = {191},
          doi = {10.1086/131977},
       adsurl = {https://ui.adsabs.harvard.edu/abs/1987PASP...99..191S}
}

@ARTICLE{2024Ap&SS.369...23H,
       author = {{H{\o}g}, Erik},
        title = "{A review of 70 years with astrometry: From meridian circles to Gaia and beyond}",
      journal = {Astrophysics and Space Science},
         year = 2024,
        month = feb,
       volume = {369},
       number = {2},
          eid = {23},
        pages = {23},
          doi = {10.1007/s10509-024-04285-8},
archivePrefix = {arXiv},
       eprint = {2402.10996},
 primaryClass = {astro-ph.IM},
       adsurl = {https://ui.adsabs.harvard.edu/abs/2024Ap&SS.369...23H}
}

@article{king1983accuracy,
  title={Accuracy of measurement of star images on a pixel array.},
  author={King, Ivan R},
  journal={Publications of the Astronomical Society of the Pacific},
  volume={95},
  number={564},
  pages={163},
  year={1983},
  publisher={IOP Publishing}
}

@article{ricker2015tess,
  author    = {Ricker, George R. and Winn, Joshua N. and Vanderspek, Roland and Latham, David W. and Bakos, G{\'a}sp{\'a}r {\'A}. and Bean, Jacob L. and Berta-Thompson, Zachory K. and Brown, Timothy M. and Buchhave, Lars A. and Butler, Nathaniel R. and et al.},
  title     = {Transiting Exoplanet Survey Satellite (TESS)},
  journal   = {Journal of Astronomical Telescopes, Instruments, and Systems},
  volume    = {1},
  number    = {1},
  pages     = {014003},
  year      = {2015},
  doi       = {10.1117/1.JATIS.1.1.014003}
}

@article{liaudat2023point,
  title={Point spread function modelling for astronomical telescopes: a review focused on weak gravitational lensing studies},
  author={Liaudat, Tob{\'\i}as I and Starck, Jean-Luc and Kilbinger, Martin},
  journal={Frontiers in Astronomy and Space Sciences},
  volume={10},
  pages={1158213},
  year={2023},
  publisher={Frontiers Media SA}
}

@article{casetti2023star,
  title={Star-image Centering with Deep Learning: HST/WFPC2 Images},
  author={Casetti-Dinescu, Dana I and Girard, Terrence M and Baena-Galle, Roberto and Martone, Max and Schwendemann, Kate},
  journal={Publications of the Astronomical Society of the Pacific},
  volume={135},
  number={1047},
  pages={054501},
  year={2023},
  publisher={IOP Publishing}
}

@article{libralato2024high,
  title={High-precision astrometry and photometry with the JWST/MIRI imager},
  author={Libralato, Mattia and Argyriou, Ioannis and Dicken, Dan and Mar{\'\i}n, Macarena Garc{\'\i}a and Guillard, Pierre and Hines, Dean C and Kavanagh, Patrick J and Kendrew, Sarah and Law, David R and Noriega-Crespo, Alberto and others},
  journal={Publications of the Astronomical Society of the Pacific},
  volume={136},
  number={3},
  pages={034502},
  year={2024},
  publisher={IOP Publishing}
}

@article{moffat1969theoretical,
  title={A theoretical investigation of focal stellar images in the photographic emulsion and application to photographic photometry},
  author={Moffat, AFJ},
  journal={Astronomy and Astrophysics, Vol. 3, p. 455 (1969)},
  volume={3},
  pages={455},
  year={1969}
}

@article{trujillo2001effects,
  title={The effects of seeing on S{\'e}rsic profiles--II. The Moffat PSF},
  author={Trujillo, I and Aguerri, JAL and Cepa, J and Guti{\'e}rrez, CM},
  journal={Monthly Notices of the Royal Astronomical Society},
  volume={328},
  number={3},
  pages={977--985},
  year={2001},
  publisher={Blackwell Science Ltd Oxford, UK}
}

@article{xin2018study,
  title={A Study of the Point-spread Function in SDSS Images},
  author={Xin, Bo and Ivezi{\'c}, {\v{Z}}eljko and Lupton, Robert H and Peterson, John R and Yoachim, Peter and Jones, R Lynne and Claver, Charles F and Angeli, George},
  journal={The Astronomical Journal},
  volume={156},
  number={5},
  pages={222},
  year={2018},
  publisher={IOP Publishing}
}

@article{nardiello2022photometry,
  title={Photometry and astrometry with JWST--I. NIRCam point spread functions and the first JWST colour--magnitude diagrams of a globular cluster},
  author={Nardiello, D and Bedin, LR and Burgasser, A and Salaris, M and Cassisi, Santi and Griggio, M and Scalco, M},
  journal={Monthly Notices of the Royal Astronomical Society},
  volume={517},
  number={1},
  pages={484--497},
  year={2022},
  publisher={Oxford University Press}
}

@article{schmitz2020euclid,
  title={Euclid: Nonparametric point spread function field recovery through interpolation on a graph Laplacian},
  author={Schmitz, Morgan A and Starck, J-L and Mboula, F Ngole and Auricchio, N and Brinchmann, Jarle and Capobianco, RI Vito and Cl{\'e}dassou, Rodolphe and Conversi, L and Corcione, L and Fourmanoit, N and others},
  journal={Astronomy \& Astrophysics},
  volume={636},
  pages={A78},
  year={2020},
  publisher={EDP Sciences}
}

@article{nie2021point,
  title={The point spread function reconstruction--II. The smooth PCA},
  author={Nie, Lin and Li, Guoliang and Peterson, John R and Wei, Chengliang},
  journal={Monthly Notices of the Royal Astronomical Society},
  volume={503},
  number={3},
  pages={4436--4445},
  year={2021},
  publisher={Oxford University Press}
}

@article{stone2023astrophot,
  title={astrophot: fitting everything everywhere all at once in astronomical images},
  author={Stone, Connor J and Courteau, St{\'e}phane and Cuillandre, Jean-Charles and Hezaveh, Yashar and Perreault-Levasseur, Laurence and Arora, Nikhil},
  journal={Monthly Notices of the Royal Astronomical Society},
  volume={525},
  number={4},
  pages={6377--6393},
  year={2023},
  publisher={Oxford University Press}
}

@article{rowell2021gaia,
  title={Gaia Early Data Release 3-Modelling and calibration of Gaia’s point and line spread functions},
  author={Rowell, N and Davidson, M and Lindegren, L and Van Leeuwen, F and Casta{\~n}eda, J and Fabricius, C and Bastian, U and Hambly, NC and Hern{\'a}ndez, J and Bombrun, A and others},
  journal={Astronomy \& Astrophysics},
  volume={649},
  pages={A11},
  year={2021},
  publisher={EDP Sciences}
}

@article{lindegren2018gaia,
  title={Gaia data release 2-the astrometric solution},
  author={Lindegren, Lennart and Hern{\'a}ndez, J and Bombrun, A and Klioner, S and Bastian, U and Ramos-Lerate, M and De Torres, A and Steidelm{\"u}ller, H and Stephenson, C and Hobbs, D and others},
  journal={Astronomy \& astrophysics},
  volume={616},
  pages={A2},
  year={2018},
  publisher={EDP Sciences}
}

@article{espinosa2025impact,
  title={The impact of the point spread function fitting radius on photometric uncertainty based on the Fisher information matrix},
  author={Espinosa, Sebastian and Vicu{\~n}a, Mario L and Mendez, Rene A and Silva, Jorge F and Orchard, Marcos},
  journal={Astronomy \& Astrophysics},
  volume={702},
  pages={A91},
  year={2025},
  publisher={EDP Sciences}
}

@article{stubbs2007toward,
  title={Toward more precise survey photometry for PanSTARRS and LSST: measuring directly the optical transmission spectrum of the atmosphere},
  author={Stubbs, Christopher W and High, F William and George, Matthew R and DeRose, Kimberly L and Blondin, St{\'e}phane and Tonry, John L and Chambers, Kenneth C and Granett, Benjamin R and Burke, David L and Smith, R Chris},
  journal={Publications of the Astronomical Society of the Pacific},
  volume={119},
  number={860},
  pages={1163},
  year={2007},
  publisher={IOP Publishing}
}

@article{becker2007pursuit,
  title={In pursuit of LSST science requirements: A comparison of photometry algorithms},
  author={Becker, Andrew C and Silvestri, Nicole M and Owen, Russell E and Ivezi{\'c}, {\v{Z}}eljko and Lupton, Robert H},
  journal={Publications of the Astronomical Society of the Pacific},
  volume={119},
  number={862},
  pages={1462},
  year={2007},
  publisher={IOP Publishing}
}

@article{gai2022astrometric,
  title={Astrometric precision tests on TESS data},
  author={Gai, Mario and Vecchiato, Alberto and Riva, Alberto and Busonero, Deborah and Lattanzi, Mario and Bucciarelli, Beatrice and Crosta, Mariateresa and Qi, Zhaoxiang},
  journal={Publications of the Astronomical Society of the Pacific},
  volume={134},
  number={1033},
  pages={035004},
  year={2022},
  publisher={IOP Publishing}
}

@article{prod2012impact,
  title={The impact of CCD radiation damage on Gaia astrometry--I. Image location estimation in the presence of radiation damage},
  author={Prod'homme, Thibaut and Holl, Berry and Lindegren, Lennart and Brown, Anthony GA},
  journal={Monthly Notices of the Royal Astronomical Society},
  volume={419},
  number={4},
  pages={2995--3017},
  year={2012},
  publisher={Oxford University Press}
}

@article{crowley2016gaia,
  title={Gaia data release 1-On-orbit performance of the Gaia CCDs at L2},
  author={Crowley, Cian and Kohley, Ralf and Hambly, Nigel C and Davidson, M and Abreu, A and Van Leeuwen, F and Fabricius, Claus and Seabroke, George and de Bruijne, JHJ and Short, A and others},
  journal={Astronomy \& Astrophysics},
  volume={595},
  pages={A6},
  year={2016},
  publisher={EDP Sciences}
}



\end{document}